\documentclass[12pt]{article}

\usepackage{newtxtext,newtxmath}

\usepackage{graphicx}

\usepackage[letterpaper,margin=1in]{geometry}

\renewenvironment{abstract}
	{\quotation}
	{\endquotation}

\date{}

\makeatletter
\renewcommand{\fnum@figure}{\textbf{Figure \thefigure}}
\renewcommand{\fnum@table}{\textbf{Table \thetable}}
\makeatother

\usepackage{scicite}

\usepackage{url}

\def\scititle{
Physically Consistent Global Atmospheric Data Assimilation with Machine Learning in Latent Space
}
\title{\bfseries \boldmath \scititle}

\author{
        Hang~Fan$^{1,2,3,4}$, 
        Lei~Bai$^{1\ast}$, 
        Ben~Fei$^{1,5\ast}$, 
        Yi~Xiao$^{1}$, 
        Kun~Chen$^{1}$, \\ 
        Yubao~Liu$^{4\ast}$, 
        Yongquan~Qu$^{2,3}$, 
        Fenghua~Ling$^{1}$, 
        Pierre~Gentine$^{2,3}$\and
	\small$^{1}$Shanghai Artificial Intelligence Laboratory, Shanghai, China.\and
	\small$^{2}$Department of Earth and Environmental Engineering, Columbia University, New York, NY, USA.\and
        \small$^{3}$Learning the Earth with Artificial intelligence and Physics (LEAP) Center, \and 
        \small Columbia University, New York, NY, USA.\and
        \small$^{4}$Nanjing University of Information Science and Technology, Nanjing, China.\and
        \small$^{5}$The Chinese University of Hong Kong, Hong Kong, China.\and
	\small$^\ast$Corresponding author. Email: bailei@pjlab.org.cn; benfei@cuhk.edu.hk; ybliu@nuist.edu.cn\and
}

\begin{document} 

\maketitle

\begin{abstract} \bfseries \boldmath
Data assimilation (DA) integrates observations with model forecasts to produce optimized atmospheric states, whose physical consistency is critical for stable weather forecasting and reliable climate research. 
Traditional Bayesian DA methods enforce these nonlinear, flow-dependent physical constraints through empirical and tunable covariance structures, but with limited accuracy and robustness.
Here, we introduce Latent Data Assimilation (LDA), a framework that performs Bayesian DA in a latent space learned from multivariate global atmospheric data via an autoencoder.
We demonstrate that the autoencoder can largely capture nonlinear physical relationships, enabling LDA to produce balanced analyses without explicitly modeling physical constraints.
Assimilation in latent space also improves both analysis quality and forecast skill compared to traditional model-space DA, under both idealized and real observational settings. 
Furthermore, LDA exhibits strong robustness across latent dimensions and remains effective even when the autoencoder is trained on inaccurate but physically realistic forecasts, highlighting its flexibility for real-world applications.

\end{abstract}

\noindent
Data assimilation (DA) aims to estimate the optimal state of dynamical systems by integrating all available information~\cite{kalnayAtmosphericModelingData2002, evensenDataAssimilationFundamentals2022}. In atmospheric science, DA is indispensable for both initializing numerical weather prediction (NWP) and generating reliable climate reanalysis datasets (e.g., the fifth-generation ECMWF reanalysis, ERA5)~\cite{carrassiDataAssimilationGeosciences2018}.
Despite significant advances in the quantity and quality of observations, contributing to a ``quiet revolution'' in forecasting~\cite{bauerQuietRevolutionNumerical2015}, the core DA methodologies have remained largely unchanged over the past two decades. This persistence can be attributed to the rigorous Bayesian design of state-of-the-art DA methods, such as four-dimensional variational assimilation (4DVar)~\cite{courtierStrategyOperationalImplementation1994} and the Ensemble Kalman Filter (EnKF)~\cite{evensenSequentialDataAssimilation1994,houtekamerDataAssimilationUsing1998}, which integrate multi-source information—including observations, model forecasts (background fields), uncertainties, model dynamics, and ensemble statistics—in a statistically optimal manner~\cite{bannisterReviewOperationalMethods2017, carrassiDataAssimilationGeosciences2018}.

Despite their theoretical rigor, current Bayesian DA methods face fundamental limitations that constrain further improvements. 
A central challenge lies in reliably estimating the background error covariance matrix (\textbf{B}), which captures the underlying physical relationships among atmospheric variables (Fig.~\ref{fig:fig1}) and is crucial for ensuring physically consistent analyses.~\cite{kalnayAtmosphericModelingData2002,bannisterReviewForecastError2008, barkerThreeDimensionalVariationalData2004, bannisterReviewOperationalMethods2017}. 
However, constructing an accurate $\mathbf{B}$ matrix remains highly challenging for two reasons: 1) the high dimensionality of NWP model states typically leads to $\mathbf{B}$ with dimensions exceeding $10^{12}$, and 2) the nonlinearity of atmospheric dynamics requires \textbf{B} to evolve with time, making it strongly flow-dependent~\cite{bannisterReviewForecastError2008, bannisterReviewOperationalMethods2017}.
Consequently, significant efforts have been devoted to estimating, simplifying, and adjusting the $\mathbf{B}$ matrix—a process that is typically empirical and involves many assumptions and approximations~\cite{parrishNationalMeteorologicalCenters1992,houtekamerDataAssimilationUsing1998, houtekamerReviewEnsembleKalman2016,hamillHybridEnsembleKalman2000a,wangHybridETKF3DVAR2008a}. 
Nevertheless, these approaches still struggle to produce a sufficiently accurate $\mathbf{B}$ matrix, leading to an imbalanced solution space in DA and an inability to fully capture the atmosphere’s inherent nonlinear structures.

Machine learning (ML), known for its ability to learn complex nonlinear mappings from high-dimensional and extensive datasets, has become an increasingly powerful tool in scientific research, including DA~\cite{geerLearningEarthSystem2021, dubenMachineLearningECMWF2021,chengMachineLearningData2023a}. 
Recent studies have leveraged ML to assist traditional Bayesian DA frameworks—by improving observation quality~\cite{wangDeepLearningAugmented2022,vandalGlobalAtmosphericData2024}, reducing computational costs~\cite{maulikEfficientHighdimensionalVariational2022,chattopadhyayDeepLearningenhancedEnsemblebased2023,xiaoFengWu4DVarCouplingDatadriven2023,liFuXiEn4DVarAssimilationSystem}, or post-processing analysis outputs~\cite{ruckstuhlTrainingConvolutionalNeural2021,farchiUsingMachineLearning2021}—while preserving the fundamental assimilation algorithms.
Beyond these developments, several new DA paradigms based on ML have also emerged. These include 1) diffusion-based methods~\cite{rozetScorebasedDataAssimilation, huangDiffDADiffusionModel2024,quDeepGenerativeData2024, manshausenGenerativeDataAssimilation2024}, which generate atmospheric analyses via observation-guided denoising, and 2) end-to-end schemes~\cite{chenEndtoendArtificialIntelligence2024, xuFuxiDAGeneralizedDeep2024, sunFuXiWeatherDatatoforecast2024, xiangADAFArtificialIntelligence2024} , which treat reanalysis data as truth and learn a direct mapping from real observations and background states to analysis states. 
Although proven effective, these approaches still lack the rigorous incorporation of prior information as comprehensively as traditional Bayesian DA (e.g., various background uncertainties, observation uncertainty, and model dynamics), thereby limiting their ability to handle complex operational assimilation scenarios. 
In addition, physical constraints among atmospheric variables are learned only implicitly in these two approaches and remain difficult to enforce explicitly. These limitations highlight the importance of developing DA frameworks that can ideally combine the non-linear representational capacity of ML with the statistical rigor of traditional methods.

Latent Data Assimilation (LDA) represents a promising avenue to address these challenges ~\cite{peyronLatentSpaceData2021, amendolaDataAssimilationLatent2021, chengMultidomainEncoderDecoder2024, melinc3DVarDataAssimilation2024, fanNovelLatentSpace2025, fanNovelLatentSpace2025a}.
Rather than operating in the original model space, LDA performs assimilation in a compact latent space learned via nonlinear encoding of high-dimensional atmospheric states using autoencoders (AEs)~\cite{hintonReducingDimensionalityData2006a} or variational autoencoders (VAEs)~\cite{kingmaAutoEncodingVariationalBayes2022}. 
By optimizing the latent space to preserve maximal information from the input atmospheric states, LDA can potentially capture essential nonlinear dependencies among variables. This property may reduce the reliance on $\mathbf{B}$ matrix while still maintaining physical consistency in analyses.
While LDA has shown promise in early studies, three critical questions remain unresolved. 
First, theoretical foundations of LDA are not yet fully established, especially regarding the underlying mechanisms that contribute to its effectiveness.
Second, existing applications have been restricted to univariate or idealized systems~\cite{maulikEfficientHighdimensionalVariational2022, melinc3DVarDataAssimilation2024, fanNovelLatentSpace2025, fanNovelLatentSpace2025a}, leaving its scalability to realistic, multivariate atmospheric problems unverified.
Third, operational feasibility for real-world assimilation and forecasting has yet to be demonstrated.

In this study, we present the first practical implementation of LDA to a high-dimensional, multivariate global atmospheric setting, and investigate the factors underlying its effectiveness.
We demonstrate that the latent representation, learned via nonlinear compression, enables physically consistent DA, with a near diagonal background error covariance matrix, which provides an advantage compared to the challenging task of defining an optimal background covariance matrix in model space. 
By analyzing the properties of latent-space increments, we show that LDA can approximate traditional DA in model space, while in practice it outperforms traditional methods in both analysis and forecasts, and is easier to implement. 
Finally, we investigate the impact of latent dimensionality on assimilation performance, providing practical guidelines for selecting an optimal latent space for LDA.

\section*{Results}
\begin{figure}[p]
    \centering
    \includegraphics[width=\linewidth]{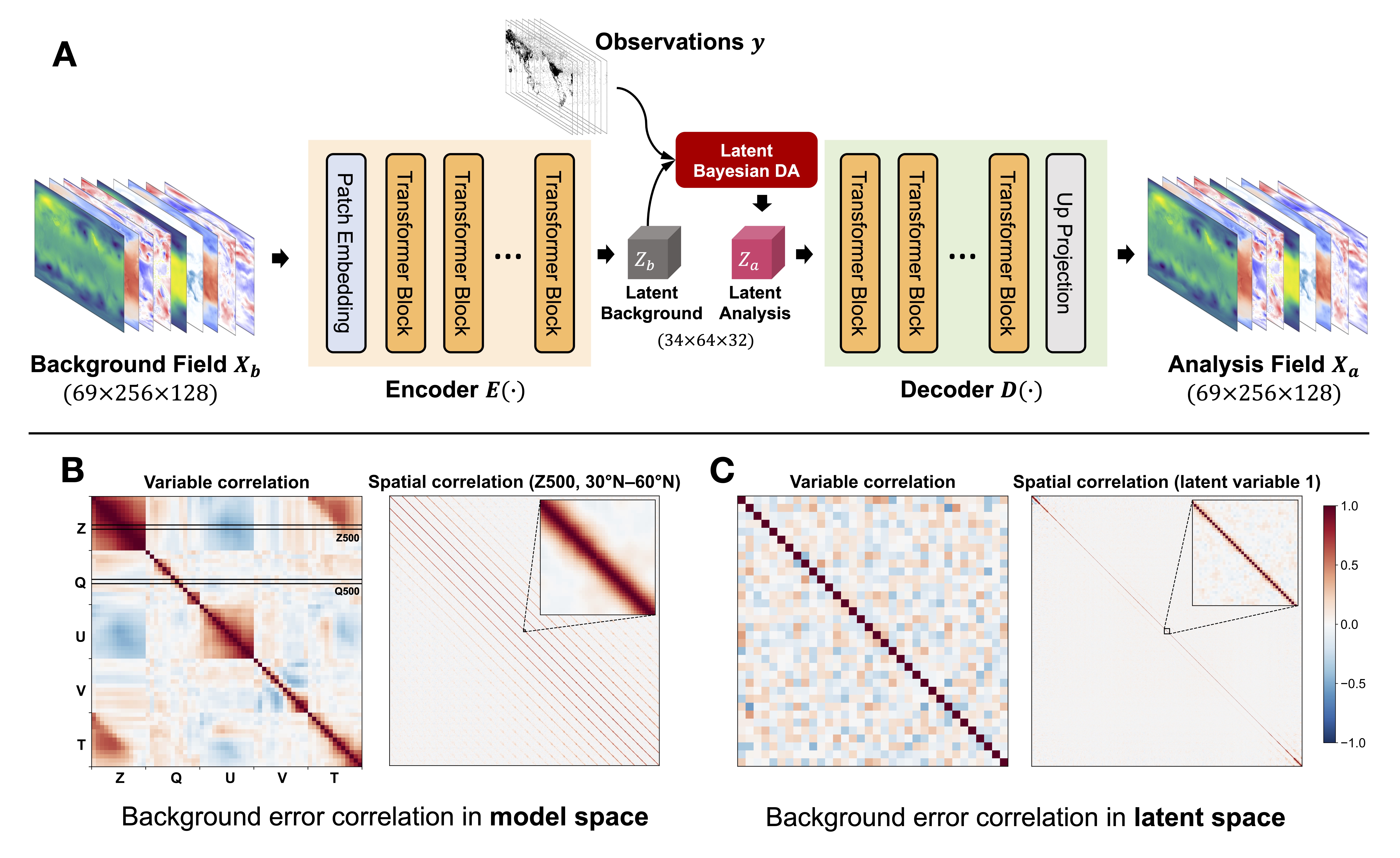}
    \caption{\textbf{Architecture of the LDA framework and comparison of background error correlations in model and latent spaces.} (\textbf{A}), Illustration of LDA for global atmosphere. The high-dimensional background atmospheric state $\boldsymbol{x_b}$ is encoded into a compact latent representation $\boldsymbol{z_b}$, here via a Swin Transformer-based autoencoder. A Bayesian variational assimilation is then performed in the latent space using observations $\boldsymbol{y}$, yielding a latent analysis $\boldsymbol{z_a}$, which is decoded to produce the analysis state $\boldsymbol{x_a}$.
    (\textbf{B--C}), Background error correlations in the model space (\textbf{B}) and latent space (\textbf{C}) estimated by ``NMC'' method. 
    The model space exhibits strong and complex inter-variable and long-range spatial correlations, while the latent space shows near-diagonal structures in both variable and spatial dimensions, indicating the decorrelation effect induced by the AE. The correlations of Z and Q at 500 hPa with other variables are highlighted by black lines in \textbf{A}.
    }
    \label{fig:fig1}
\end{figure}

We implement LDA by performing Bayesian assimilation in a low-dimensional representation of the atmospheric state,  learned via an AE with an encoder–decoder architecture (Fig.~\ref{fig:fig1}A). 
Specifically, the encoder first compresses the high-dimensional atmospheric background fields into a latent representation, where observations are then assimilated; the decoder subsequently reconstructs the latent analysis to produce the final model-space analysis. 
We train the AE on ERA5 reanalysis~\cite{hersbachERA5GlobalReanalysis2020} at 1.40625° resolution, covering 69 atmospheric variables—four at the surface and five in the upper air across 13 pressure levels. The model compresses the full atmospheric state (69×256×128) into a latent space of size (34×64×32), preserving both variable structure and spatial organization. 

To implement a four-dimensional (4D) LDA and evaluate its ability to support accurate weather forecasts, we train an end-to-end forecast model using the same ERA5 dataset employed for AE training. The model architecture follows FengWu~\cite{chenFengWuPushingSkillful2023}, a Swin Transformer–based medium-range AI weather prediction system. 
While such end-to-end models exhibit strong forecasting skill~\cite{biAccurateMediumrangeGlobal2023, lamLearningSkillfulMediumrange2023, chenFengWuPushingSkillful2023, chenFuXiCascadeMachine2023}, their limited ability to propagate small initial perturbations and accurately represent the ensemble spread has been noted in previous studies~\cite{selzCanArtificialIntelligenceBased2023, slivinskiAssimilatingObservedSurface}. Accordingly, we demonstrate the superiority of LDA by comparing the variational DA in model and latent spaces, using a static background error covariance estimated via the National Meteorological Center (NMC) method~\cite{parrishNationalMeteorologicalCenters1992}. For clarity, we denote the variational DA methods as ``3DVar'' and ``4DVar'' when applied in model space, and as ``L3DVar'' and ``L4DVar'' when implemented in latent space. Implementation details of the AE, forecast model, and the DA methods are provided in the Methods section.

In variational DA, obtaining the optimal analysis requires inverting the $\mathbf{B}$ matrix, which is computationally prohibitive due to the high-dimensional nature of weather forecasting.
This matrix in model space typically exhibits strong spatial and cross-variable correlations (Fig.~\ref{fig:fig1}B) and is therefore often simplified into a diagonal form via empirical control variable transform~\cite{descombesGeneralizedBackgroundError2015} to facilitate inversion. 
Within the LDA framework, although the latent space can be substantially lower-dimensional than the model space, a full-rank latent background error covariance matrix ($\mathbf{B}_z$) still contains up to $10^9$ elements, making direct inversion impractical. 
Fortunately, consistent with previous findings from single-variable LDA studies~\cite{melinc3DVarDataAssimilation2024,fanNovelLatentSpace2025a}, we find that $\mathbf{B}_z$ is naturally approximately diagonal, with uniformly small off-diagonal elements (Fig.~\ref{fig:fig1}C). 
This property allows for a simple yet accurate approximation of $\mathbf{B}_z^{-1}$ by taking the reciprocals of its diagonal entries. Consequently, all subsequent LDA experiments adopt this diagonalized form of $\mathbf{B}_z$.

\subsection*{Superiority of assimilation in latent space over model space}
We first evaluate the effectiveness of LDA using observing system simulation experiments (OSSEs)~\cite{arnoldObservingSystemsSimulationExperiments1986,atlasAtmosphericObservationsExperiments1997,hoffmanFutureObservingSystem2016}, a standard framework for benchmarking DA methods under controlled and idealized conditions. 
In these experiments, the ERA5 reanalysis is treated as the true atmospheric state for validation, from which synthetic observations are sampled based on the location of the GDAS (Global Data Assimilation System). 
In each experiment, observations from four time steps at six-hour intervals are assimilated into a 54-hour forecast initialized from the ERA5 reanalysis, after which a 10-day free forecast is performed.

\begin{figure}[p]
    \centering
    \includegraphics[width=\linewidth]{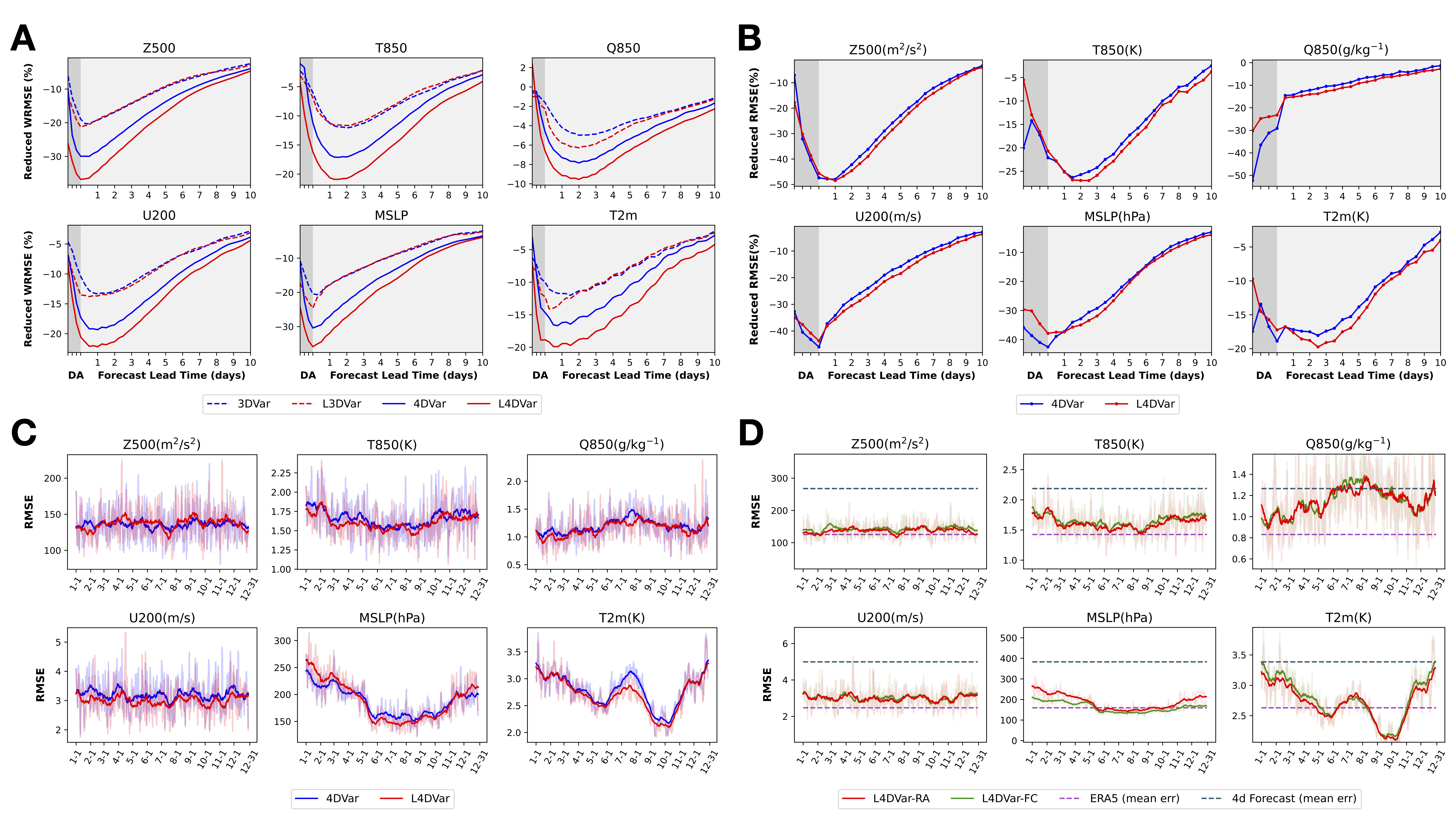}
    \caption{\textbf{Performance comparison of Variational DA methods in model space and latent space across OSSEs and real-world experiments.} 
    (\textbf{A}), Percentage reduction in RMSE relative to a control run without DA in idealized OSSEs, evaluated against ERA5. Results are averaged over daily experiments in 2017, each initialized at 0000 UTC and assimilating four time steps at 6-hour intervals. Observations were sampled from ERA5 using the radiosonde and surface observation locations of GDAS at 0000 UTC on January 1.  
    (\textbf{B}), Same as \textbf{A}, but using real GDAS observations in 2017, which include radiosonde and surface data available at 12-hour intervals. Accuracy is evaluated against all available observations.  
    (\textbf{C}), analysis RMSE from L4DVar and 4DVar using GDAS observations over the full year of 2017, computed against 16 radiosonde and 300 surface observations withheld from assimilation. Shaded lines indicate daily variability; solid lines show smoothed trends.  
    (\textbf{D}), Comparison of L4DVar performance using AEs trained on ERA5 reanalysis (L4DVar-RA) and 5-day forecasts (L4DVar-FC). Average errors of the respective training datasets are shown as dashed lines.}
    \label{fig:fig2}
\end{figure}

Fig.~\ref{fig:fig2}A shows the mean percentage reduction in analysis and forecast errors for each DA method, relative to a control run without assimilation, averaged over daily OSSEs conducted throughout 2017. 
Overall, the variational DA methods perform substantially better in the latent space than in the model space, for both analyses and forecasts. In particular, this advantage is particularly pronounced in L4DVar, which reduces the analysis error by an average of 5.1\% relative to 4DVar and maintains its superiority throughout the forecast period for most variables. In contrast, L3DVar achieves a 3\% reduction in analysis error compared to 3DVar and shows reduced improvement in forecast persistence. These results indicate that the model dynamics can be effectively utilized and play a significant role in LDA.
 
Beyond idealized OSSEs, we further evaluate the performance of LDA using real surface and radiosonde observations from GDAS throughout 2017. 
After quality control and interpolation (see Methods for details), the processed dataset includes over 400 upper-air and 3,000 surface observations with an interval of 12 hours.
Fig.~\ref{fig:fig2}B evaluates 10-day forecasts initialized from L4DVar and 4DVar analyses against all conventional observations collected by GDAS. Both methods improve forecast skill, with L4DVar consistently outperforming 4DVar, across nearly all variables.
Interestingly, although L4DVar and 4DVar produce noticeably different analyses at the end of assimilation, their forecast errors become statistically similar in the early hours—in contrast with OSSEs, where the advantage of L4DVar progressively amplifies during the initial forecast steps.
A likely explanation is that LDA analyses, derived from real observations, deviate from the ERA5 distribution used to train the ML-based forecast model, leading to a model adjustment that temporarily narrows the performance gap.

To evaluate the analysis capability of LDA, we conducted cycling assimilation experiments throughout 2017, with a 48-hour DA window comprising four observation steps. The accuracy of the resulting analysis fields from L4DVar and 4DVar was validated using observations withheld from assimilation (Fig. \ref{fig:fig2}C).
Across all pressure levels and time steps, L4DVar produces more accurate analyses than 4DVar for 54 of the 69 variables, with a mean reduction in error approximately 5\%.
These results highlight the practical superiority of LDA in both idealized and real-world settings, supporting its potential as a more effective alternative to traditional DA frameworks.

Since the AE in LDA is trained in an unsupervised manner, we conduct a proof-of-concept experiment to examine whether LDA remains effective when the AE is trained on less accurate data. Specifically, we compare the LDA performance using an AE trained on ERA5 reanalysis versus one trained on 4-day forecasts (Fig.~\ref{fig:fig2}D). Remarkably, despite the fact that the forecast data exhibiting substantially larger errors than ERA5, L4DVar based on the forecast-trained AE produces reanalyses of comparable quality to those based on ERA5—and still outperforms traditional 4DVar. More importantly, LDA yields reanalyses that are significantly more accurate than the data used for training. These results suggest that, given sufficient observations, LDA can enhance the accuracy of analysis products beyond the limitations of the training data. An expanded version of Fig.~\ref{fig:fig2}, covering a more comprehensive set of variables across multiple vertical levels, is provided in figs. S1–4.

\subsection*{Physical consistency of LDA}
\begin{figure}[p]
    \centering
    \includegraphics[width=\linewidth]{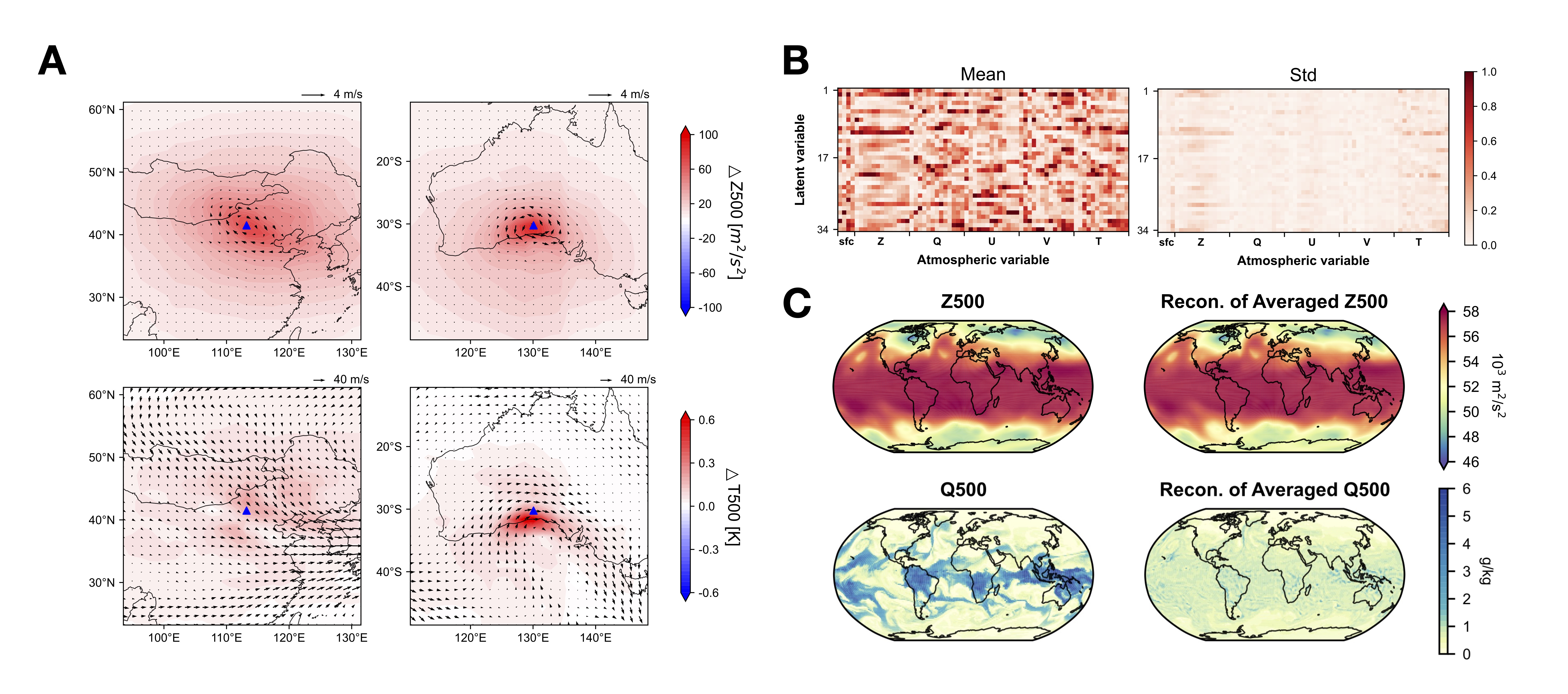}
    \caption{\textbf{Evidence of physical consistency in latent-space assimilation.} 
    (\textbf{A}), Analysis increments from a $+200 m^2/s^2$ perturbation applied to Z500 at 200\,hPa over China (left) and Australia (right), using ERA5 reanalysis at 0000 UTC on 1 December 2017 as the background. Top panels show geopotential height and wind increments; bottom panels show temperature increments and the background wind field. Perturbation locations are marked with blue triangles.
    (\textbf{B}), Influence of latent variables on atmospheric variables. Shown are the mean and standard deviation of perturbations in atmospheric variables induced by perturbations in each latent variable, computed over 10,000 randomly selected pairs from the ERA5 test dataset. In each sample, the model-space perturbation is spatially averaged and normalized.
    (\textbf{C}), AE reconstruction of physically imbalanced inputs. The fields of Z500 and Q500 at 0000 UTC on 1 January 2017 are first averaged globally and then passed through the AE. }
    \label{fig:fig3}
\end{figure}

In model-space DA, physical balance among variables is implicitly enforced through the covariance structure of $\mathbf{B}$, which becomes nearly diagonal in latent space. 
To evaluate whether LDA maintains physical consistency under such a diagonal $\mathbf{B}_z$, we conducted single-observation experiments with L3DVar. 
Fig.~\ref{fig:fig3}A shows the analysis increment resulting from a $+200 m^2/s^2$ geopotential height perturbation at 500 hPa, centered over China (left) and Australia (right). 
The geopotential height increment decays radially from the perturbation center, inducing anticyclonic wind anomalies.
Notably, the wind response reverses direction between hemispheres, consistent with geostrophic balance at 500 hPa. 
The lower panel of Fig.~\ref{fig:fig3}A displays the associated temperature increment, which closely follows the background wind flow. 
Fig. S5 presents the same observation increment as in Fig.~\ref{fig:fig3}A, but under a different background state, leading to a distinct analysis increment.
These results show that, even with a diagonal and static $\mathbf{B}_z$, LDA can produce physically consistent and flow-dependent analyses through the state-varying encoding, suggesting that physical constraints is implicitly encoded in the latent space.

These findings imply that the latent representation captures the inter-variable dependencies directly, making it unnecessary to rely on a complex $\mathbf{B}_z$ covariance to enforce physical constraints. Instead those constraints are learned through the encoding process.
To verify this hypothesis, we first analyze the impact of each latent variable (i.e., the first dimension of the latent state) on atmospheric variables.
As shown in Fig.~\ref{fig:fig3}B, each latent variable consistently influences specific atmospheric variables to varying degrees, indicating that the latent representation captures inter-variable dependencies critical for maintaining physical consistency.
Fig. S6 shows that different latent variables exert approximately orthogonal influences, revealing the decorrelation effect of the encoding. 

To further investigate the inherent physical constraints of the latent representation, we test the ability of the AE to reconstruct physically unbalanced atmospheric states. Specifically, we replace one variable in an ERA5 sample with its spatial mean and evaluate the reconstruction (Fig.~\ref{fig:fig3}C). The results indicate that the averaged Z500 field is accurately recovered, while Q500 is not, reflecting differences in their dependency on other variables. 
As shown in Fig.~\ref{fig:fig1}B, the geopotential height (Z500), which exhibits strong vertical and inter-variable correlations, can be largely reconstructed even from an unphysical averaged input. In contrast, the specific humidity (Q500), being more independent, proves much harder to reconstruct.

\subsection*{Why LDA works?}
Although LDA achieves promising results in both analysis and forecast accuracy, its underlying mechanism remains unclear, raising potential questions about its overall reliability. 
A central challenge stems from the fact that LDA solves the optimal analysis in the latent space, without guaranteeing that decoding the optimal latent state will yield an optimal analysis in model space as well. Mathematically, we demonstrate that LDA and traditional model-space assimilation are equivalent if and only if the decoder $D(\cdot)$ is locally affine and error-free throughout the assimilation process, under an assumption of zero decoding error (see Supplementary Text).

This theoretical insight motivates us to evaluate the local behavior of the decoder along latent space increments $\Delta \boldsymbol{z}$ that captures leading modes of atmospheric variability, which dominate the structure of DA updates.
To this end, we construct $\Delta \boldsymbol{z}$ by differencing latent states from randomly selected ERA5 reanalysis samples, and examine their effects on the decoded outputs. Extensive experiments reveal two key properties: the decoder response is approximately additive with respect to latent increments and nearly homogeneous in magnitude, as presented in Fig.~\ref{fig:fig4}A. Additional examples and statistical results provided in the Supplementary Materials (fig. S8 and tabel S1) indicate that the decoder behaves approximately affine near each latent state $\boldsymbol{z}$ along directions of  $\Delta \boldsymbol{z}$. 
This behavior can be explained by a first-order Taylor expansion: a continuous function, including highly nonlinear neural networks, is approximately affine within sufficiently small regions. 

To estimate the valid range of the decoder linearity, we test whether scaled latent increments $k \cdot \Delta \boldsymbol{z}$ produce proportionally scaled outputs. 
As shown in Fig.~\ref{fig:fig4}B, the decoder exhibits approximate affine behavior across a broad range of $k$ along latent directions $\Delta \boldsymbol{z}$ that represent dominant atmospheric variability. 
Given that DA analysis increments in model space reflect similar underlying variability but are much smaller in magnitude, the corresponding latent increments in LDA are expected to align with the directions of the tested $\Delta \boldsymbol{z}$ and remain within the decoder’s locally affine regime. 
Fig. S9 supports this by showing that in both OSSE and real-observation experiments (Fig.~\ref{fig:fig2}A, B), the resulting analysis increments follow directions where the decoder is locally affine around the background state. These findings suggest why LDA can effectively and stably approximate the model-space optimal analysis in this study.

\subsection*{Impact of Latent Dimensionality on LDA}
The dimensionality of the latent space is a key parameter that influences the LDA performance. To investigate its effects, we evaluate both analysis and forecast skill using AEs trained with varying compression ratios.
As the latent space is three-dimensional in this study, we evaluate multiple spatial compression ratios $C_{\text{spatial}}$ under each fixed total compression ratio $C_{\text{total}}$. 
As shown in Fig.~\ref{fig:fig5}A–C, LDA consistently outperforms traditional DA across a wide range of compression ratios, demonstrating strong robustness. L4DVar, which incorporates model dynamics, exhibits notably lower sensitivity to compression ratios than L3DVar. 
Interestingly, as $C_{\text{total}}$ increases, LDA performance initially improves and then declines, regardless of $C_{\text{spatial}}$, with the optimal value consistently being around 32, across all experimental settings.

The existence of an optimal latent size in LDA reflects a trade-off between two competing effects induced by AE compression. On the one hand, it can increase the reconstruction error due to information loss during encoding (Fig.~\ref{fig:fig5}D), which may degrade the analysis quality. On the other hand, compression enhances the decorrelation of latent variables, supporting the validity of diagonalizing $\mathbf{B}_z$. 
This effect is illustrated in fig. S10, where increasing $C_{\text{spatial}}$ reduces spatial correlations in $\mathbf{B}_z$. 
As the LDA performance initially improves with increasing $C_{\text{total}}$ (Fig.~\ref{fig:fig5}A-C), this suggests that the decorrelation effect dominates at first.
However, since this benefit saturates (fig. S4), the reconstruction error grows continuously. 
As a result, an optimal latent size always exists and can be efficiently identified via binary search~\cite{cormen2022introduction}.

We note that a patch-like increment sometimes emerges in single-observation experiments when the latent variable dimension exceeds that of the model space (fig. S11), likely due to the ViT-based architecture of the AE. 
While this effect is mitigated with dense observations and remains acceptable in our experiments (Fig.~\ref{fig:fig5}A–C), we recommend compressing both variable and spatial dimensions when using ViT-based AEs. Accordingly, we adopt a latent size of 34×64×32 in this study.

\section*{Discussion}
Advances in DA rely on the integration of new sources of information, yet few methodological breakthroughs have emerged in the past two decades. 
The rapid development of ML and the increasing availability of atmospheric data offer a new opportunity to revolutionize DA. In this study, we construct a latent representation of the global atmosphere using an AE trained on decades of ERA5 reanalysis and perform DA directly in this latent space. We demonstrate that LDA enables physically constrained analyses and holds great potential for improving both analysis and forecasting tasks.

While this study demonstrates that Bayesian DA in latent space can approximate the accuracy its counterpart in the model space, our experiments indicate that latent-space assimilation offers superior practical performance. This improvement can be attributed to their fundamental differences in the treatment of atmospheric inter-variable correlations. Traditional DA frameworks depend on the quality of the background covariance $\textbf{B}$ to define a physically constrained solution space~\cite{bannisterReviewForecastError2008}. However, accurately estimating $\textbf{B}$ remains extremely challenging due to its high dimensionality and flow-dependent characteristics. In contrast, the latent space learned by the AE naturally encodes these non-linear physical relationships through the compression process. As a result, in LDA, $\textbf{B}_z$ in LDA primarily characterizes the magnitude of background uncertainty rather than enforcing physical consistency, and can therefore be effectively diagonalised. This enables LDA to be a simpler, more physically consistent, and more effective alternative to traditional model-space DA approaches. It is worth noting that the $\textbf{B}_z$ used in this study retains some spatial and cross-variable correlations. Decomposing these components as in traditional variational DA~\cite{descombesGeneralizedBackgroundError2015} may further improve LDA performance.

The effectiveness of the LDA relies on two empirical properties: the approximate affinity of the decoder in DA process and the near-diagonal structure of $\textbf{B}_z$. While these properties are difficult to guarantee, similar findings have been reported across LDA studies involving different variables, scales, and AE architectures~\cite{melinc3DVarDataAssimilation2024, zhengGeneratingUnseenNonlinear,fanNovelLatentSpace2025a}, suggesting that such behaviors may be general under certain conditions.
A possible explanation for the decoder’s local affinity is that assimilation increments are typically small compared to climatological variability, allowing the decoder to be well-approximated by a first-order Taylor expansion along directions that represent atmospheric variability.
The near-diagonal structure of $\textbf{B}_z$ likely arises from the AE’s compression objective, which promotes decoupling among atmospheric variables, as evidenced by the latent impact analysis in fig. S6.
These characteristics underpin the potential of LDA to generalize to more complex Earth system models, including high-resolution atmospheric and oceanic simulations.

Many existing end-to-end DA methods are trained to reproduce the reanalysis used as labels\cite{chenEndtoendArtificialIntelligence2024, sunFuXiWeatherEndtoend2024} and therefore struggle to exceed its accuracy, regardless of the amount of observations assimilated. In contrast, LDA employs an unsupervised AE that only requires high-fidelity input data with physically consistent atmospheric structures, without relying on accurate analyses or real observations. 
Even when the AE is trained on forecasts with substantial errors, LDA can still produce reliable analyses—only slightly less accurate than those based on reanalysis-trained models and significantly outperforming its own training data. This suggests that, with more assimilated observations, LDA has the potential to surpass existing reanalysis products such as ERA5~\cite{hersbachERA5GlobalReanalysis2020}, which are based on 4DVar.
Moreover, this capability is especially valuable in data-sparse or underdeveloped regions, where long-term reanalysis is difficult and expensive to construct, but physically consistent mesoscale forecasts initialized from global analyses remain accessible.

While LDA demonstrates notable advantages over both traditional and ML-based DA methods, its effectiveness remains limited by several factors, including the AE reconstruction error, the Gaussian assumption of background errors in latent space, and the architecture of the AE network. These challenges underscore the need for continued advancements in AE design to further improve assimilation performance. In addition, the current implementation of L4DVar requires a differentiable forecasting model, limiting its applicability to ML–based systems. As physics-informed neural models and end-to-end forecasting frameworks continue to advance—and increasingly demonstrate potential to replace traditional CPU-based systems\cite{biAccurateMediumrangeGlobal2023,lamLearningSkillfulMediumrange2023, kochkovNeuralGeneralCirculation2024}—this constraint is expected to diminish. We believe LDA provides a promising pathway toward next-generation DA systems, with the potential for integration into high-resolution, hybrid Earth system models.

\section*{Materials and Methods}\label{sec:method}
\subsection*{Reanalysis dataset, forecast model, and observations}\label{sec4.1}
The AE, forecast model, and OSSE experiments are all based on ERA5 reanalysis data~\cite{hersbachERA5GlobalReanalysis2020}, with a spatial resolution of 1.4° latitude/longitude encompassing a global grid of 128 × 256 points.  
This dataset comprises 69 variables: four surface variables and five upper-air atmospheric variables across 13 pressure levels (50, 100, 150, 200, 250, 300, 400, 500, 600, 700, 850, 925, and 1000 hPa). The surface variables included 2-meter air temperature (t2m), 10-meter wind (u10, v10), and mean sea-level pressure (msl), while the upper-air variables included geopotential height (Z), temperature (T), zonal wind (U), meridional wind (V), and specific humidity (Q). 
ERA5 reanalysis data from 1979 to 2015 were used to train the AE and forecast model, with 2016 reserved for validation and 2017 for testing and data assimilation experiments. The dataset was used at hourly resolution for AE training and 6-hourly resolution for the forecast model.

Our ML-based forecast model follows the architecture of FengWu~\cite{chenFengWuPushingSkillful2023}, which employs a modality-specific design with separate Swin Transformer~\cite{liuSwinTransformerHierarchical2021} encoders and decoders for each atmospheric variable, and fuses cross-variable information via a combination of Vision Transformer and Swin Transformer blocks. Unlike the original FengWu model, which takes atmospheric states from two consecutive time steps as input, our implementation uses only a single time step to support 4DVar assimilation.
Furthermore, since our study involves cyclic DA to generate analysis and requires high-fidelity forecast data to train the AE, we do not apply multi-step autoregressive fine-tuning~\cite{chenFuXiCascadeMachine2023,lamLearningSkillfulMediumrange2023}, which may smooth out the forecasts.

The global observations for the assimilation experiments are sourced from GDAS in 2017. As a preliminary investigation, only surface and radiosonde observations with the BUFR codes of ``ADPUPA'' and ``ADPSFC'' are utilized. 
All observations are interpolated onto the model state grid for simplicity, and redundant observations falling within the same grid point are averaged to reduce sampling noise.
High-elevation surface observations are treated as upper-air observations after altitude interpolation. 
In real-observation assimilation experiments, we discard observations whose deviations from ERA5 analyses exceed variable-specific thresholds to ensure assimilation stability. These thresholds are defined as the mean absolute difference between ERA5 analyses separated by 48 hours, computed over the year 2016 for all 69 variables. To account for representativeness errors and interpolation uncertainties associated with terrain effects, thresholds for surface observations are inflated by a factor of four.
Post-processing finally yielded over 3000 surface and 400 radiosonde observations every 12 hours.

\subsection*{AE for compressing multivariate global atmosphere}\label{sec4.2}
The autoencoder (AE)~\cite{hintonReducingDimensionalityData2006a} comprises an encoder \( E(\cdot) \) and a decoder \( D(\cdot) \) (Fig.~\ref{fig:fig1}A). The encoder maps the multivariate global atmospheric state \( \boldsymbol{x} \) in model space to a latent representation \( \boldsymbol{z} \), while the decoder reconstructs \( \boldsymbol{x} \) from \( \boldsymbol{z} \). The AE is trained by minimizing the mean squared reconstruction error,$\left\| \boldsymbol{x} - D(E(\boldsymbol{x})) \right\|_2,$ ensuring that the latent representation \( \boldsymbol{z} = E(\boldsymbol{x}) \) preserves the essential information in \( \boldsymbol{x} \).

Our AE architecture is built upon a vision transformer with window attention~\cite{liuSwinTransformerHierarchical2021}, following the design presented in~\cite{hanCRA5ExtremeCompression2024} for compressing the ERA5 dataset.

The encoder begins with a 4×4 patch embedding that partitions the input field into 64×32 tokens, each represented by a 1024-dimensional vector, followed by 12 Swin Transformer blocks with 16 attention heads.
Subsequently, a shared fully connected network (FCN) is applied to each token to project them into a latent space of shape 34×64×32.
The decoder mirrors the encoder structure but applies a FCN to each latent token after the transformer blocks, followed by reshaping to recover the model-space structure.
The model was trained for 30 epochs using the AdamW optimizer~\cite{loshchilovDecoupledWeightDecay2019}. 
A learning rate of $2 \times 10^{-4}$ was employed, incorporating a linear warm-up followed by a cosine decay schedule.

\subsection*{Traditional variational DA methods in model space}\label{sec4.3}

Assuming the errors of the background field should be singular $\boldsymbol{x_b}$ and observations $\boldsymbol{y}$ are Gaussian and independent, the maximum posterior estimate of the atmospheric state $\boldsymbol{x}$ can be obtained by minimizing the variational cost function $J(\boldsymbol{x})$. 
For 3DVar~\cite{courtierECMWFImplementationThreedimensional1998}, the cost function is as follows:
\begin{equation}
J(\boldsymbol{x})=\frac{1}{2}(\boldsymbol{x}-\boldsymbol{x_b})^\mathrm{T} \textbf{B} ^{-1}(\boldsymbol{x}-\boldsymbol{x_b}) + \frac{1}{2} (\boldsymbol{y}-\mathcal{H}(\boldsymbol{x}))^\mathrm{T} \textbf{R}^{-1} (\boldsymbol{y}-\mathcal{H}(\boldsymbol{x})),
\end{equation}
where, $\textbf{B}$ and $\textbf{R}$ represent the error covariance matrix for $\boldsymbol{x_b}$ and $\boldsymbol{y}$, respectively.
$\mathcal{H}(\cdot)$ denotes the observation operator, facilitating a projection from model space to the observational space of $\boldsymbol{y}$. 

4DVar simultaneously assimilates a sequence of observations over a time window by incorporating model dynamics as a constraint~\cite{courtierStrategyOperationalImplementation1994}. In this study, we adopt a strongly constrained 4DVar formulation, which is also the standard approach used in operational systems such as the Integrated Forecasting System (IFS)~\cite{ecmwfIFSDocumentationCY49R12024}. The cost function is defined as:
\begin{equation}
J(\boldsymbol{x})=\frac{1}{2}(\boldsymbol{x}-\boldsymbol{x_b})^\mathrm{T} \textbf{B}^{-1}(\boldsymbol{x}-\boldsymbol{x_b}) + \frac{1}{2} \sum_{i=0}^{n} (\boldsymbol{y_i}-\mathcal{H}(M_{0\rightarrow i}(\boldsymbol{x})))^\mathrm{T} \textbf{R}_i^{-1} (\boldsymbol{y_i}-\mathcal{H}(M_{0\rightarrow i}(\boldsymbol{x}))),
\end{equation}
where the subscript $i=0,1,\ldots,n$ denotes sequential time points, and $M_{0\rightarrow i}$ represents model forecast operator from the initial time to $t_i$. 

Traditional 3DVar and 4DVar minimization employ an iterative gradient descent approach, requiring the computation of $\nabla J(\boldsymbol{x})$. 
This is particularly challenging for 4DVar, necessitating programming tangent linear and adjoint models~\cite{bannisterReviewOperationalMethods2017}. 
Fortunately, the use of neural network–based models facilitates automated minimization of $J(\boldsymbol{x})$ via gradient descent and backpropagation~\cite{lecunGradientbasedLearningApplied1998}. 
This process resembles neural network training but uses $J(\boldsymbol{x})$ as the loss function and optimizes only the model states $\boldsymbol{x}$. 
We employ the L-BFGS optimizer for 3DVar due to its efficiency. 
However, for 4DVar, the incorporation of a nonlinear ML model makes its cost function not strictly convex, necessitating the use of a stochastic optimizer like Adam~\cite{kingmaAdamMethodStochastic2017}.

\subsection*{The variational LDA methods}\label{sec4.4}

In LDA, the latent background state \( \boldsymbol{z}_b \) is defined as the encoded representation of \( \boldsymbol{x}_b \), obtained via the AE.
Denoting the error covariance matrix of $\boldsymbol{z_b}$ as $\textbf{B}_z$, the cost function of 3DVar in latent space can be expressed as:
\begin{equation}
J(\boldsymbol{z})=\frac{1}{2}(\boldsymbol{z}-\boldsymbol{z_b})^\mathrm{T} \textbf{B}_z ^{-1}(\boldsymbol{z}-\boldsymbol{z_b}) + \frac{1}{2} (\boldsymbol{y}-\mathcal{H}(D(\boldsymbol{z})))^\mathrm{T} \textbf{R}^{-1} (\boldsymbol{y}-\mathcal{H}(D(\boldsymbol{z}))),
\end{equation}
and for L4Dvar, it is formulated as follows:
\begin{equation}
J(\boldsymbol{z})=\frac{1}{2}(\boldsymbol{z}-\boldsymbol{z_b})^\mathrm{T} \textbf{B}_z^{-1}(\boldsymbol{z}-\boldsymbol{z_b}) + \frac{1}{2} \sum_{i=0}^{n} (\boldsymbol{y_i}-\mathcal{H}(M_{0\rightarrow i}(D(\boldsymbol{z})))^\mathrm{T} \textbf{R}_i^{-1} (\boldsymbol{y_i}- \mathcal{H}(M_{0\rightarrow i}(D(\boldsymbol{z})))).
\end{equation}
Minimizing these two functions yields the latent-space analysis \( \boldsymbol{z}_a \), whose decoded counterpart \( \boldsymbol{x}_a = D(\boldsymbol{z}_a) \) represents the LDA analysis in model space.
Note that the minimization process requires stochastic optimization techniques due to the incorporation of the nonlinear decoder.

\subsection*{Estimation of \textbf{B} matrix}\label{sec4.5}
The classical NMC method \cite{parrishNationalMeteorologicalCenters1992} is employed to provide the static forecast error covariance matrix of forecast model. 
The $\textbf{B}$-matrix in model space is estimated as follows:
\begin{equation}
\textbf{B}\approx  \frac{1}{2} \left \langle  (\boldsymbol{x}^{48}-\boldsymbol{x}^{24})(\boldsymbol{x}^{48}-\boldsymbol{x}^{24})^\mathrm{T}  \right \rangle,
\end{equation}
where $\boldsymbol{x}^{24}$ and $\boldsymbol{x}^{48}$ represent the 48 h and 24 h forecasts valid at the same time, and $\left \langle\cdot\right\rangle$ denotes the average over a large number of samples. 
To address the computational and storage challenges posed by $\textbf{B}$ matrix, we followed the NCAR-developed GEN\_BE 2.0 method \cite{descombesGeneralizedBackgroundError2015}, which generates background error matrices for the WRF model. 
This method decomposes $\textbf{B}$ matrix into several components: $\textbf{B}=\textbf{U}\textbf{U}^\mathrm{T}$, where $\textbf{U}=\textbf{U}_p\textbf{S}\textbf{U}_v\textbf{U}_h$. The $\textbf{U}_p$, $\textbf{U}_v$, $\textbf{U}_h$,  $\textbf{S}$ matrix represents the physical variable correlation, vertical correlation, horizontal correlation, and the diagonal standard deviations of the decomposed variables, respectively.

In the latent space, the computation of the background error covariance matrix \( \mathbf{B}_z \) becomes substantially simpler, as it is approximately diagonal by construction. 
Specifically, we calculate each diagonal element of $\textbf{B}_z$ with the NMC method as follows:
\begin{equation}
\mathbf{B}_{z,i} \approx \frac{1}{2} \left\langle \left( E(\boldsymbol{x}^{48})_i - E(\boldsymbol{x}^{24})_i \right)^2 \right\rangle,
\end{equation}
where $i$ denotes the $i_{th}$ element of the latent space variable, and $E(\cdot)$ represents the AE encoder.

We computed $\boldsymbol{x}^{24}$ and $\boldsymbol{x}^{48}$ at 6-hourly intervals throughout 2016, yielding 1460 paired samples, to provide $\textbf{B}$ and $\textbf{B}_z$ required for DA experiments in 2017. 
Note that both \( \mathbf{B} \) and \( \mathbf{B}_z \) are derived from the same forecast samples, ensuring a consistent environment for comparing DA and LDA. The length scales of \( \mathbf{B} \) and the magnitude of \( \mathbf{B}_z \) were tuned through experiments conducted in 2016.

\subsection*{Metrics}\label{sec4.6}

In OSSEs, given the ground truth is available at each grid point,
we utilize the latitude-weighted root mean square error (WRMSE) to quantify the error of each atmospheric variable $c$ of the model field $x$ as follows:
\begin{equation}
\operatorname{WRMSE}(\boldsymbol{x},\boldsymbol{x}_{truth}, c)=\sqrt{\frac{1}{H \cdot W} \sum_{h, w} H \cdot \frac{\cos \left(\alpha_{h, w}\right)}{\sum_{h^{\prime}=1}^{H} \cos \left(\alpha_{h^{\prime}, w}\right)}\left(\boldsymbol{x}^{c,h,w}-\boldsymbol{x}_{truth}^{c,h,w}\right)^{2}},
\end{equation}
where superscript $c, h, w$ denote the index for variables, latitude grid, and longitude grid, respectively.
$\alpha_{h, w}$ is the latitude of point $(h,w)$. 
$H$ and $W$ represent the number of grid points in the longitudinal and latitudinal directions of the model space.

For real observation experiments, since the observation used for validation is sparse, we assess the accuracy of $\boldsymbol{x}$ by directly calculating the root mean square error (RMSE) of each atmospheric variable at the observation locations as follows:
\begin{equation}
\operatorname{RMSE}(\boldsymbol{x}, \boldsymbol{y}, c)=\sqrt{\frac{1}{N} \sum_{i} (\mathcal{H}(\boldsymbol{x})^{c,i}-\boldsymbol{y}^{c,i})^{2}},
\end{equation}
where the superscript $c,i$ denote the $i_{th}$ observation of variable $c$, and $N$ represents the total number of observations for that variable.



\clearpage

\clearpage

\clearpage

\clearpage
\begin{figure}[p]
    \centering
    \includegraphics[width=\linewidth]{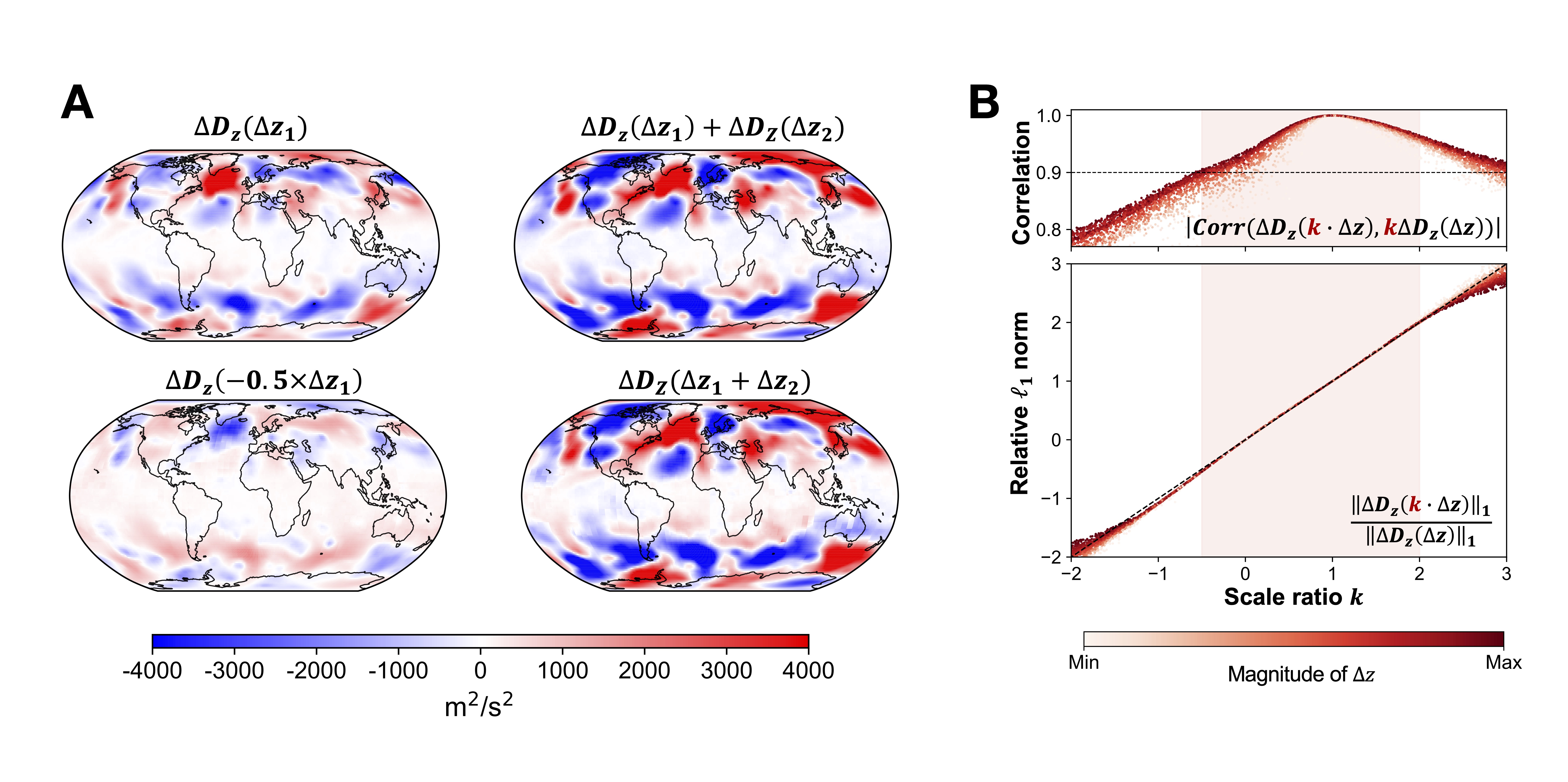}
    \caption{\textbf{Approximate affinity of the AE decoder along latent directions representative of atmospheric variability.} 
    (\textbf{A}), The impact of latent-space perturbations $\Delta \boldsymbol{z}$ at $\boldsymbol{z}$ on decoding results, denoted as $\Delta D_{\boldsymbol{z}}(\Delta \boldsymbol{z})$, shown using Z500. Here, $\boldsymbol{z}$ denotes the latent state corresponding to the ERA5 reanalysis at 0000 UTC on February 1, 2017. The perturbations $\Delta \boldsymbol{z_1}$ and $\Delta \boldsymbol{z_2}$ represent the latent differences between $\boldsymbol{z}$ and the reanalysis at 0000 UTC on January 1 and March 1, 2017, respectively. 
    (\textbf{B}), Evaluation of the near-linear response region of the AE decoder. The upper panel shows the correlation between $\Delta D_{\boldsymbol{z}}(k \cdot \Delta \boldsymbol{z})$ and $k \cdot \Delta D_{\boldsymbol{z}}(\Delta \boldsymbol{z})$ as a function of scale ratio $k$; the lower panel shows their relative $\ell_1$ norm. 
    }
    \label{fig:fig4}
\end{figure}

\clearpage
\begin{figure}[p]
    \centering
    \includegraphics[width=\linewidth]{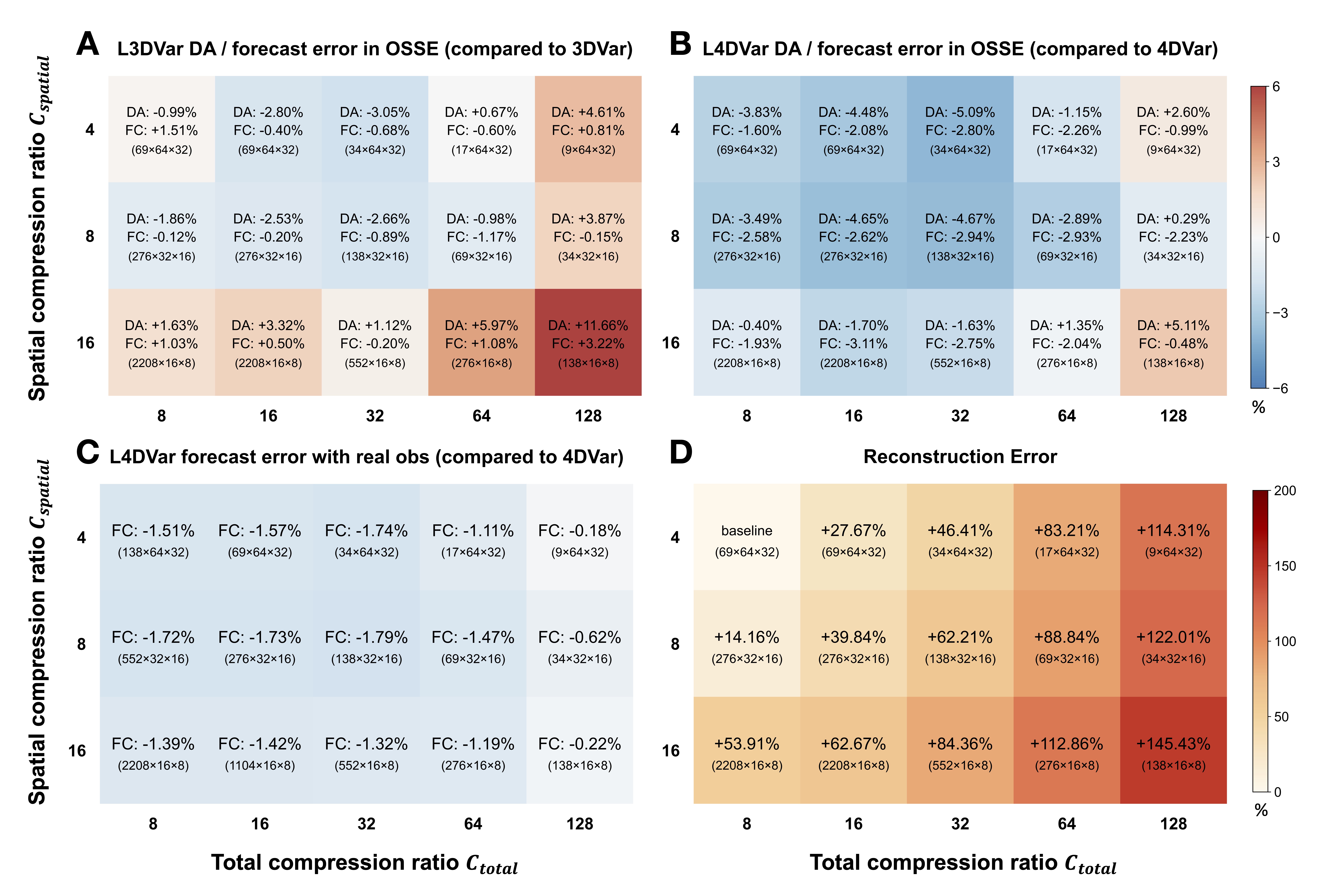}
    \caption{\textbf{Impact of the latent size on LDA performance.} 
    (\textbf{A}), Relative changes in analysis (DA) and forecast (FC) error for L3DVar compared to 3DVar in OSSEs. Experimental configuration is identical to that in Fig.~\ref{fig:fig2}A. Colors indicate the mean improvement across DA and FC. 
    (\textbf{B}), Same as (a), but for L4DVar compared to 4DVar. 
    (\textbf{C}), Forecast error changes for L4DVar with real GDAS observations, relative to 4DVar. The experimental configuration is identical to that in Fig.~\ref{fig:fig2}B.
    (\textbf{D}), AE reconstruction error for different latent sizes, expressed as percentage increase in RMSE relative to the largest latent size (138×64×32).
    }
    \label{fig:fig5}
\end{figure}




	


\clearpage 

%
\bibliography{science_template} 
\bibliographystyle{sciencemag}

%
%
%
%
%
%


\section*{Acknowledgments}
We would like to express our sincere gratitude to the ECMWF for providing the invaluable ERA5 reanalysis dataset. We also gratefully acknowledge the NCEI for making the global observations available.
\paragraph*{Funding:}
P.G. and Y.Q. acknowledge support from the National Science Foundation (NSF) Science and Technology Center (STC) Learning the Earth with Artificial Intelligence and Physics (LEAP), Award \#2019625 and USMILE European Research Council grant.

\paragraph*{Author contributions:}
Conceptualization: H.F., Y.L and P.G. Methodology: H.F. Data curation: H.F. and K.C. Software: H.F., Y.X., and B.F. Formal analysis: H.F., Y.X, Y.Q, and F.L. Writing – original draft: H.F., L.B., B.F., and P.G. Writing – review and editing: All authors. Supervision: L.B. and P.G. Visualization: H.F. Resources: L.B.

\paragraph*{Competing interests:}
There are no competing interests to declare.

\paragraph*{Data and materials availability:}
The ERA5 dataset is available from the official website of Climate Data Store (CDS) at \url{https://cds.climate.copernicus.eu/.} The GDAS observational BUFR files can be accessed via the NCEI Archive Information Request System (AIRS) at \url{https://www.ncei.noaa.gov/has/HAS.DsSelect}. The neural network model is developed using PyTorch. Codes and model checkpoints used in this study are available at \url{https://github.com/hangfan99/LDA_1.41}.


\subsection*{Supplementary materials}
Supplementary Text\\
Figs. S1 to S11\\
Tables S1


\newpage


\renewcommand{\thefigure}{S\arabic{figure}}
\renewcommand{\thetable}{S\arabic{table}}
\renewcommand{\theequation}{S\arabic{equation}}
\renewcommand{\thepage}{S\arabic{page}}
\setcounter{figure}{0}
\setcounter{table}{0}
\setcounter{equation}{0}
\setcounter{page}{1} 


\begin{center}
\section*{Supplementary Materials for\\ \scititle}

Hang~Fan$^{1,2,3,4}$, 
Lei~Bai$^{1\ast}$, 
Ben~Fei$^{1,5\ast}$, 
Yi~Xiao$^{1}$, \\
Kun~Chen$^{1}$, 
Yubao~Liu$^{4\ast}$, 
Yongquan~Qu$^{2,3}$, 
Fenghua~Ling$^{1}$, 
Pierre~Gentine$^{2,3}$\\
\small$^\ast$Corresponding author. Email: bailei@pjlab.org.cn; benfei@cuhk.edu.hk; ybliu@nuist.edu.cn\\
\end{center}

\subsubsection*{This PDF file includes:}
Supplementary Text\\
Figs. S1 to S11\\
Tables S1

\newpage


\subsection*{Supplementary Text}
\subsubsection*{Derivation of variational assimilation methods in latent space}
We begin by deriving the latent-space 3DVar (L3DVar) cost function using a Bayesian maximum a posteriori (MAP) estimation framework. Specifically, given a latent background estimate \( \boldsymbol{z}_b \) and observations \( \boldsymbol{y} \), L3DVar seeks the latent state $\boldsymbol{z}$ that maximizes the posterior distribution $p(\boldsymbol{z} \mid \boldsymbol{y},\boldsymbol{z}_b )$. Since the observations are usually independent with the background, the posterior distribution is proportional to the product of the likelihood and the prior:
\begin{equation}
p(\boldsymbol{z} \mid \boldsymbol{y},\boldsymbol{z}_b ) \propto p(\boldsymbol{y} \mid \boldsymbol{z}) \, p(\boldsymbol{z} \mid \boldsymbol{z}_b).
\end{equation}

Following the derivation of 3DVar, we assume a Gaussian prior distribution for the latent variable \( \boldsymbol{z} \), centered at the latent background state \( \boldsymbol{z}_b \) with covariance \( \mathbf{B}_z \):
\begin{equation}
p(\boldsymbol{z} \mid \boldsymbol{z}_b) = \mathcal{N}(\boldsymbol{z}_b, \mathbf{B}_z).
\end{equation}

To verify the rationality of this assumption, we analyzed a large sample of background errors generated using the NMC method. As shown in Fig. \ref{fig:Gaussian_distribution}, the distribution of background errors projected onto the latent space is unimodal and symmetrical, supporting the use of a Gaussian prior as a reasonable modeling choice.

For the likelihood, we recall that the observations are related to the latent variable through a composition of the decoder \( \mathcal{D} \) and the observation operator \( \mathcal{H} \), i.e., \( \boldsymbol{x} = \mathcal{D}(\boldsymbol{z}) \), followed by \( \mathcal{H}(\boldsymbol{x}) \). Assuming Gaussian observational errors with mean zero and covariance \( \mathbf{R} \), the likelihood becomes:
\begin{equation}
p(\boldsymbol{y} \mid \boldsymbol{z}) = \mathcal{N}(\mathcal{H}(\mathcal{D}(\boldsymbol{z})), \mathbf{R}).
\end{equation}
Combining the prior and likelihood, the posterior takes the form:
\begin{align}
p(\boldsymbol{z} \mid \boldsymbol{y}, \boldsymbol{z}_b) \propto 
\exp\left\{ -\frac{1}{2} (\boldsymbol{z} - \boldsymbol{z}_b)^\top \mathbf{B}_z^{-1} (\boldsymbol{z} - \boldsymbol{z}_b)
- \frac{1}{2} \left[ \mathcal{H}(\mathcal{D}(\boldsymbol{z})) - \boldsymbol{y} \right]^\top \mathbf{R}^{-1} \left[ \mathcal{H}(\mathcal{D}(\boldsymbol{z})) - \boldsymbol{y} \right] \right\}.
\end{align}

Taking the negative log-likelihood (up to a constant), we obtain the L3DVar cost function:
\begin{equation}
J_z(\boldsymbol{z}) = \frac{1}{2} (\boldsymbol{z} - \boldsymbol{z}_b)^\top \mathbf{B}_z^{-1} (\boldsymbol{z} - \boldsymbol{z}_b)
+ \frac{1}{2} \left[ \mathcal{H}(\mathcal{D}(\boldsymbol{z})) - \boldsymbol{y} \right]^\top \mathbf{R}^{-1} \left[ \mathcal{H}(\mathcal{D}(\boldsymbol{z})) - \boldsymbol{y} \right].
\label{eq:L3DVar_MAP}
\end{equation}

Compared to the derivation of conventional 3DVar in model space, L3DVar differs in two aspects: 1) it assumes the prior distribution in the latent space rather than the model space, and 2) it replaces the model-state variable \( \boldsymbol{x} \) in the likelihood with the decoded output \( \mathcal{D}(\boldsymbol{z}) \). Following the same principle, the cost function for L4DVar can be derived analogously as:
\begin{equation}
J(\boldsymbol{z})=\frac{1}{2}(\boldsymbol{z}-\boldsymbol{z_b})^\mathrm{T} \textbf{B}_z^{-1}(\boldsymbol{z}-\boldsymbol{z_b}) + \frac{1}{2} \sum_{i=0}^{n} (\boldsymbol{y_i}-\mathcal{H}(M_{0\rightarrow i}(D(\boldsymbol{z})))^\mathrm{T} \textbf{R}_i^{-1} (\boldsymbol{y_i}- \mathcal{H}(M_{0\rightarrow i}(D(\boldsymbol{z})))).
\end{equation}
where the subscript $i=0,1,\ldots,n$ indexes sequential time steps, and $M_{0\rightarrow i}$ denotes the model forecast operator that propagates the state from the initial time to time  $t_i$.

\subsubsection*{Equivalence Conditions between Latent-Space and Model-Space DA}

To establish a theoretical connection between model-space and latent-space DA, we begin with a 3D variational (3DVar) framework. We formally prove the following proposition:

\vspace{0.5em}
\noindent
\textbf{Proposition.} Let \( \mathcal{D} : \mathbb{R}^n \to \mathbb{R}^m \) be an error-free decoder mapping from latent space to model space. Then, the solution of latent-space 3DVar is equivalent to that of model-space 3DVar for all valid background states and observations if and only if \( \mathcal{D} \) is affine over the region traversed during assimilation.

\vspace{1em}
\noindent
\textbf{Proof of sufficiency.}
Let \( \boldsymbol{x}_a \) and \( \boldsymbol{z}_a \) denote the optimal analyses from 3DVar and L3DVar, respectively. The first-order optimality conditions for these solutions are given by:
\begin{equation}
\mathbf{B}^{-1} (\boldsymbol{x}_a - \boldsymbol{x}_b) + 
\mathbf{H}^\mathrm{T} \mathbf{R}^{-1} \left[ \mathcal{H}(\boldsymbol{x}_a) - \boldsymbol{y} \right] = \mathbf{0},
\label{solution_3Dvar}
\end{equation}
\begin{equation}
\mathbf{B}_z^{-1} (\boldsymbol{z}_a - \boldsymbol{z}_b) + 
\mathbf{J}_{\mathcal{D}}^\mathrm{T} \mathbf{H}^\mathrm{T} \mathbf{R}^{-1} \left[ \mathcal{H}(\mathcal{D}(\boldsymbol{z}_a)) - \boldsymbol{y} \right] = \mathbf{0},
\label{solution_L3Dvar}
\end{equation}
where \( \mathbf{J}_{\mathcal{D}}\) denotes the Jacobian of the decoder evaluated at \( \boldsymbol{z}_a \), and \( \mathbf{H} \) denotes the Jacobian of the observation operator evaluated at \( \boldsymbol{x}_a \).

As the decoder is affine throughout the assimilation process, i.e., \( \mathbf{J}_{\mathcal{D}}(\boldsymbol{z}) \equiv \mathbf{J}_{\mathcal{D}} \), we have:
\begin{align}
\mathbf{J}_{\mathcal{D}} \mathbf{B}_z \mathbf{J}_{\mathcal{D}}^\mathrm{T} &= \mathbf{J}_{\mathcal{D}} \, \mathbb{E}[(\boldsymbol{z}_b - \boldsymbol{z}_t)(\boldsymbol{z}_b - \boldsymbol{z}_t)^\mathrm{T}] \, \mathbf{J}_{\mathcal{D}}^\mathrm{T} \notag \\
&= \mathbb{E}[\mathbf{J}_{\mathcal{D}}(\boldsymbol{z}_b - \boldsymbol{z}_t)(\boldsymbol{z}_b - \boldsymbol{z}_t)^\mathrm{T} \mathbf{J}_{\mathcal{D}}^\mathrm{T}] \notag \\
&= \mathbb{E}[(\mathcal{D}(\boldsymbol{z}_b) - \mathcal{D}(\boldsymbol{z}_t))(\mathcal{D}(\boldsymbol{z}_b) - \mathcal{D}(\boldsymbol{z}_t))^\mathrm{T}] \notag \\
&= \mathbb{E}[(\boldsymbol{x}_b - \boldsymbol{x}_t)(\boldsymbol{x}_b - \boldsymbol{x}_t)^\mathrm{T}] = \mathbf{B},
\label{inverse_B}
\end{align}
Left-multiplying Eq.(\ref{solution_3Dvar}) and Eq.(\ref{solution_L3Dvar}) by \( \mathbf{B} \) and \( \mathbf{J}_{\mathcal{D}}\mathbf{B}_z \), respectively, we obtain:
\begin{equation}
(\boldsymbol{x}_a - \boldsymbol{x}_b) + \mathbf{B}
\mathbf{H}^\mathrm{T} \mathbf{R}^{-1} \left[ \mathcal{H}(\boldsymbol{x}_a) - \boldsymbol{y} \right] = \mathbf{0},
\label{solution_3Dvar_v2}
\end{equation}
\begin{equation}
\mathbf{J}_{\mathcal{D}}(\boldsymbol{z}_a - \boldsymbol{z}_b) + \mathbf{J}_{\mathcal{D}}\mathbf{B}_z \mathbf{J}_{\mathcal{D}}^\mathrm{T} \mathbf{H}^\mathrm{T} \mathbf{R}^{-1} \left[ \mathcal{H}(\mathcal{D}(\boldsymbol{z}_a)) - \boldsymbol{y} \right] = \mathbf{0},
\label{solution_L3Dvar_v2}
\end{equation}
Substituting Eq.(\ref{inverse_B}), $\mathbf{J}_{\mathcal{D}}(\boldsymbol{z}_a - \boldsymbol{z}_b)=\boldsymbol{x}_a^{\mathrm{L3DVar}} - \boldsymbol{x}_b$, and $ \mathcal{D}(\boldsymbol{z}_a) = \boldsymbol{x}_a^{\mathrm{L3DVar}} $ into Eq.(\ref{solution_L3Dvar}), we obtain:
\begin{equation}
(\boldsymbol{x}_a^{\mathrm{L3DVar}} - \boldsymbol{x}_b) + \mathbf{B}
\mathbf{H}^\mathrm{T} \mathbf{R}^{-1} \left[ \mathcal{H}(\boldsymbol{x}_a^{\mathrm{L3DVar}}) - \boldsymbol{y} \right] = \mathbf{0}.
\label{solution_L3Dvar_to_modelspace}
\end{equation}
Comparing Eq.~\eqref{solution_3Dvar_v2} and Eq.~\eqref{solution_L3Dvar_to_modelspace}, we conclude that if the decoder is affine during the DA process, then L3DVar and 3DVar yield the same solution in model space.

\vspace{3em}
\noindent
\textbf{Proof of necessity}
A first-order Taylor expansion of the decoder at $\boldsymbol{x}_a$ gives:
\begin{equation}
\boldsymbol{x}_a - \boldsymbol{x} = 
\mathbf{J}_{\mathcal{D}} (\boldsymbol{z}_a - \boldsymbol{z}) + \boldsymbol{\varepsilon},
\end{equation}
where \( \boldsymbol{\varepsilon} \) denotes the higher-order residual of the Taylor expansion, which depends on both \( \boldsymbol{z}_a \) and \( \boldsymbol{z} \). Assume:
\begin{equation}
\boldsymbol{x}_a - \boldsymbol{x}_t = 
\mathbf{J}_{\mathcal{D}} (\boldsymbol{z}_a - \boldsymbol{z}_t) + \boldsymbol{\varepsilon}_1,
\end{equation}
\begin{equation}
\boldsymbol{x}_a - \boldsymbol{x}_b = 
\mathbf{J}_{\mathcal{D}} (\boldsymbol{z}_a - \boldsymbol{z}_b) + \boldsymbol{\varepsilon}_2,
\label{eq:taylor_xb}
\end{equation}
As a result,
\begin{align}
\boldsymbol{x}_t - \boldsymbol{x}_b &= 
(\boldsymbol{x}_t - \boldsymbol{x}_a) + (\boldsymbol{x}_a - \boldsymbol{x}_b) \\
&= \mathbf{J}_{\mathcal{D}} (\boldsymbol{z}_t - \boldsymbol{z}_b) + (\boldsymbol{\varepsilon}_2 - \boldsymbol{\varepsilon}_1).
\end{align}
We can then express the background covariance as:
\begin{align}
\mathbf{B} &= \mathbb{E} \left[ (\boldsymbol{x}_t - \boldsymbol{x}_b)(\boldsymbol{x}_t - \boldsymbol{x}_b)^\mathrm{T} \right] \nonumber \\
&= \mathbb{E} \left[ 
\left( \mathbf{J}_{\mathcal{D}} (\boldsymbol{z}_t - \boldsymbol{z}_b) + (\boldsymbol{\varepsilon}_2 - \boldsymbol{\varepsilon}_1) \right)
\left( \cdots \right)^\mathrm{T} 
\right] \nonumber \\
&= \mathbf{J}_{\mathcal{D}} \mathbb{E} \left[ (\boldsymbol{z}_t - \boldsymbol{z}_b)(\boldsymbol{z}_t - \boldsymbol{z}_b)^\mathrm{T} \right] \mathbf{J}_{\mathcal{D}}^\mathrm{T} \\
&\quad + \mathbb{E} [\boldsymbol{\xi}+\boldsymbol{\xi}^\mathrm{T}]  \nonumber + \mathbb{E} \left[ (\boldsymbol{\varepsilon}_2 - \boldsymbol{\varepsilon}_1)(\boldsymbol{\varepsilon}_2 - \boldsymbol{\varepsilon}_1)^\mathrm{T} \right],
\end{align}
where, $\boldsymbol{\xi}=  \mathbf{J}_{\mathcal{D}}(\boldsymbol{z}_t - \boldsymbol{z}_a)(\boldsymbol{\varepsilon}_2 - \boldsymbol{\varepsilon}_1)$. Therefore, 
\begin{equation}
\mathbf{B} = \mathbf{J}_{\mathcal{D}} \mathbf{B}_z \mathbf{J}_{\mathcal{D}}^\mathrm{T} + \mathbb{E} [\boldsymbol{\xi}+\boldsymbol{\xi}^\mathrm{T}] +
\mathbb{E} \left[ (\boldsymbol{\varepsilon}_2 - \boldsymbol{\varepsilon}_1)(\boldsymbol{\varepsilon}_2 - \boldsymbol{\varepsilon}_1)^\mathrm{T} \right].
\label{eq:difference_B}
\end{equation}

We now derive a necessary condition under which the solutions of L3DVar and 3DVar coincide. By combining Eq.(\ref{solution_3Dvar}) and Eq.(\ref{solution_L3Dvar}), we obtain the following criterion:
\begin{equation}
\mathbf{B}_z^{-1} (\boldsymbol{z}_a - \boldsymbol{z}_b)=
\mathbf{J}_{\mathcal{D}}^\mathrm{T} \mathbf{B}^{-1} (\boldsymbol{x}_a - \boldsymbol{x}_b).
\label{eq:criterion_1}
\end{equation}
Left-multiplying Eq.(\ref{eq:criterion_1}) by \( \mathbf{J}_{\mathcal{D}}\mathbf{B}_z \), we obtain:
\begin{equation}
\mathbf{J}_{\mathcal{D}}(\boldsymbol{z}_a - \boldsymbol{z}_b)=
\mathbf{J}_{\mathcal{D}}\mathbf{B}_z\mathbf{J}_{\mathcal{D}}^\mathrm{T} \mathbf{B}^{-1} (\boldsymbol{x}_a - \boldsymbol{x}_b).
\label{eq:criterion_2}
\end{equation}
Substituting Eq.(\ref{eq:taylor_xb}) and Eq.(\ref{eq:difference_B}) in Eq.(\ref{eq:criterion_2}) yields:
\begin{align}
&\boldsymbol{x}_a - \boldsymbol{x}_b - \boldsymbol{\varepsilon}_2 = \left( \mathbf{B} 
- \mathbb{E} [\boldsymbol{\xi}+\boldsymbol{\xi}^\mathrm{T}]
- \mathbb{E}[ (\boldsymbol{\varepsilon}_2 - \boldsymbol{\varepsilon}_1)
(\boldsymbol{\varepsilon}_2 - \boldsymbol{\varepsilon}_1)^\mathrm{T}] \right) \nonumber \mathbf{B}^{-1} (\boldsymbol{x}_a - \boldsymbol{x}_b)
\end{align}
After simplification, we obtain:
\begin{equation}
\boldsymbol{\varepsilon}_2=(\mathbb{E} \left[ (\boldsymbol{\varepsilon}_2 - \boldsymbol{\varepsilon}_1)(\boldsymbol{\varepsilon}_2 - \boldsymbol{\varepsilon}_1)^\mathrm{T} \right]
-\mathbb{E} [\boldsymbol{\xi}+\boldsymbol{\xi}^\mathrm{T}])\mathbf{B}^{-1} (\boldsymbol{x}_a - \boldsymbol{x}_b).
\label{eq:criterion_final}
\end{equation}

To ensure \( \mathcal{D}(\boldsymbol{z}_a) = \boldsymbol{x}_a \), Eq.\ref{eq:criterion_final} must hold for arbitrary DA scenarios. If $\mathcal{D}$ is locally affine in the DA process, the residuals $
\boldsymbol{\varepsilon}_1$ and $\boldsymbol{\varepsilon}_2$ vanish by construction, making both terms zero and satisfying the condition. Conversely, if $\mathcal{D}$ is not affine, the 
residuals do not vanish in general, and the left-hand side becomes a sum of positive semi-definite matrices that cannot cancel for arbitrary $\boldsymbol{z_a}$ and  $\boldsymbol{z_b}$. 
For neural network-based decoders, which are generally nonlinear, this condition is almost impossible to satisfy. 
Therefore, Eq.~\ref{eq:criterion_final} establishes a necessary condition for the decoder to behave affinely during the DA process, ensuring the equivalence between L3DVar and 3DVar.

\vspace{1em}
\noindent
\textbf{Extension to EnKF and 4Dvar}
In the 3DVar case, the proof primarily hinges on the transformation between the background covariance matrices in model space and latent space. Importantly, the result does not rely on the specific form of the observation term, which can be arbitrary. Therefore, this proposition naturally extends to the 4DVar framework, where the structure of the cost function remains similar except for the inclusion of time-evolving dynamics. For the Ensemble Kalman Filter (EnKF), it has been shown that the analysis update is theoretically equivalent to that of 3DVar under the linear-Gaussian assumption. As a result, the same affine condition on the decoder remains valid in the EnKF setting as well.






\begin{figure}[p]
    \centering
    \includegraphics[width=\linewidth]{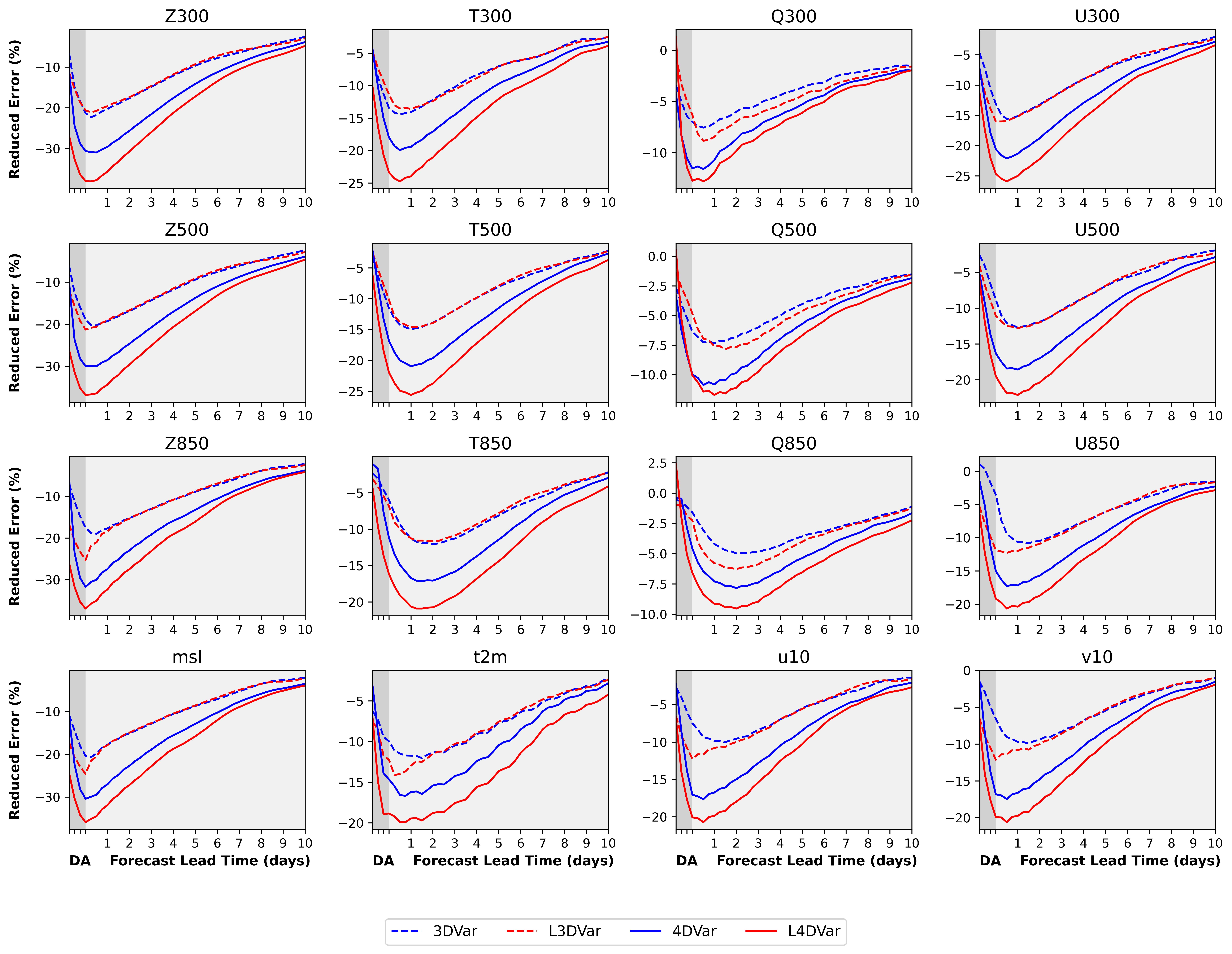}
    \caption{\textbf{Extended performance comparison of Variational DA methods in model space and latentx
    space in OSSEs.} This figure expands on Fig. 2A by showing RMSE reduction for variables at 300 hPa, 500 hPa, 850 hPa, and the surface in idealized OSSEs.}
    \label{fig:OSSE_extend}
\end{figure}

\begin{figure}[p]
    \centering
    \includegraphics[width=\linewidth]{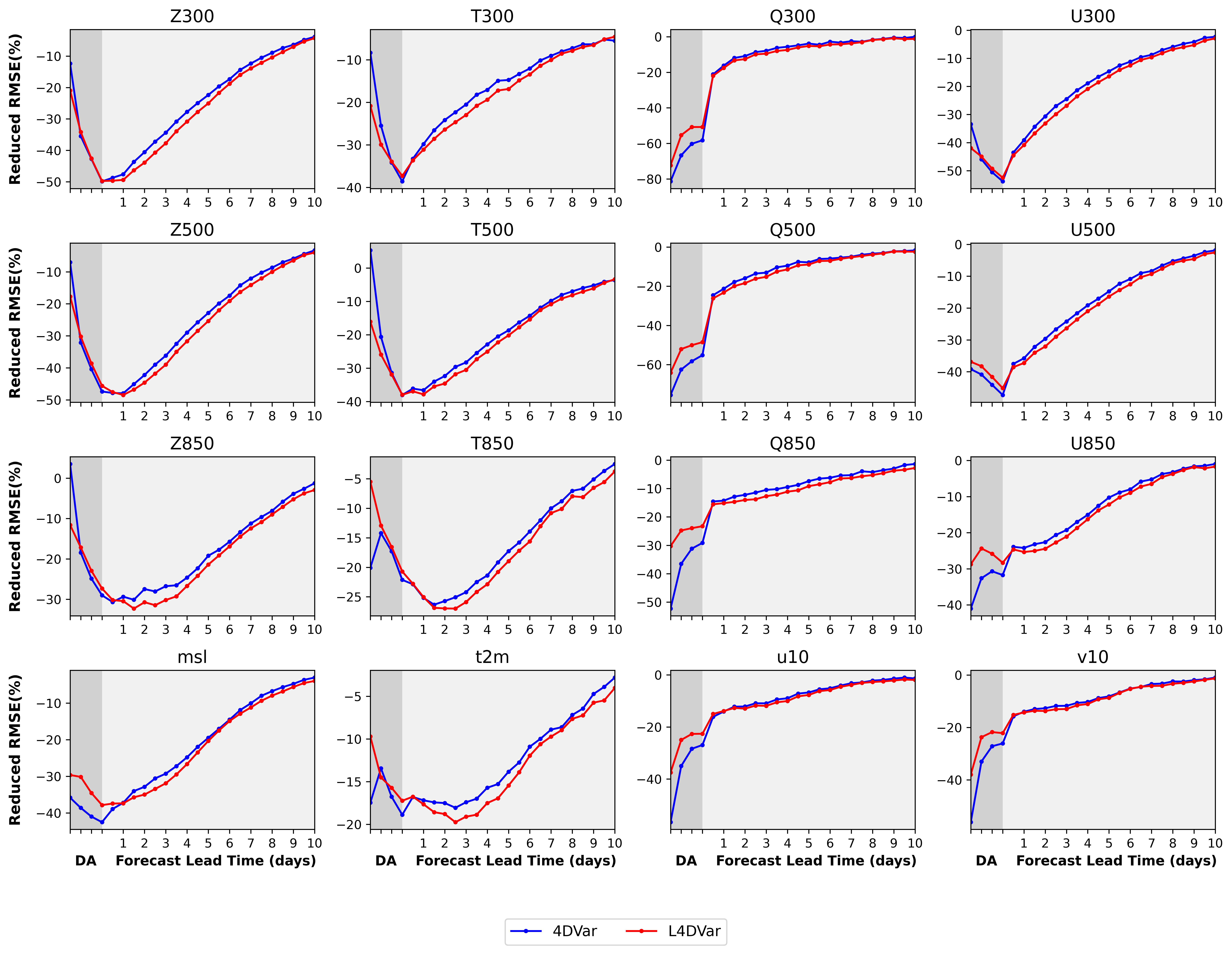}
    \caption{\textbf{Extended forecast performance comparison of 4DVar and L4DVar with real observations.} This figure expands on Fig. 2B by presenting RMSE reduction over a 10-day forecast for individual variables at 300 hPa, 500 hPa, 850 hPa, and the surface in real-observation experiments.}
    \label{fig:real_forecast_extend}
\end{figure}

\begin{figure}[p]
    \centering
    \includegraphics[width=\linewidth]{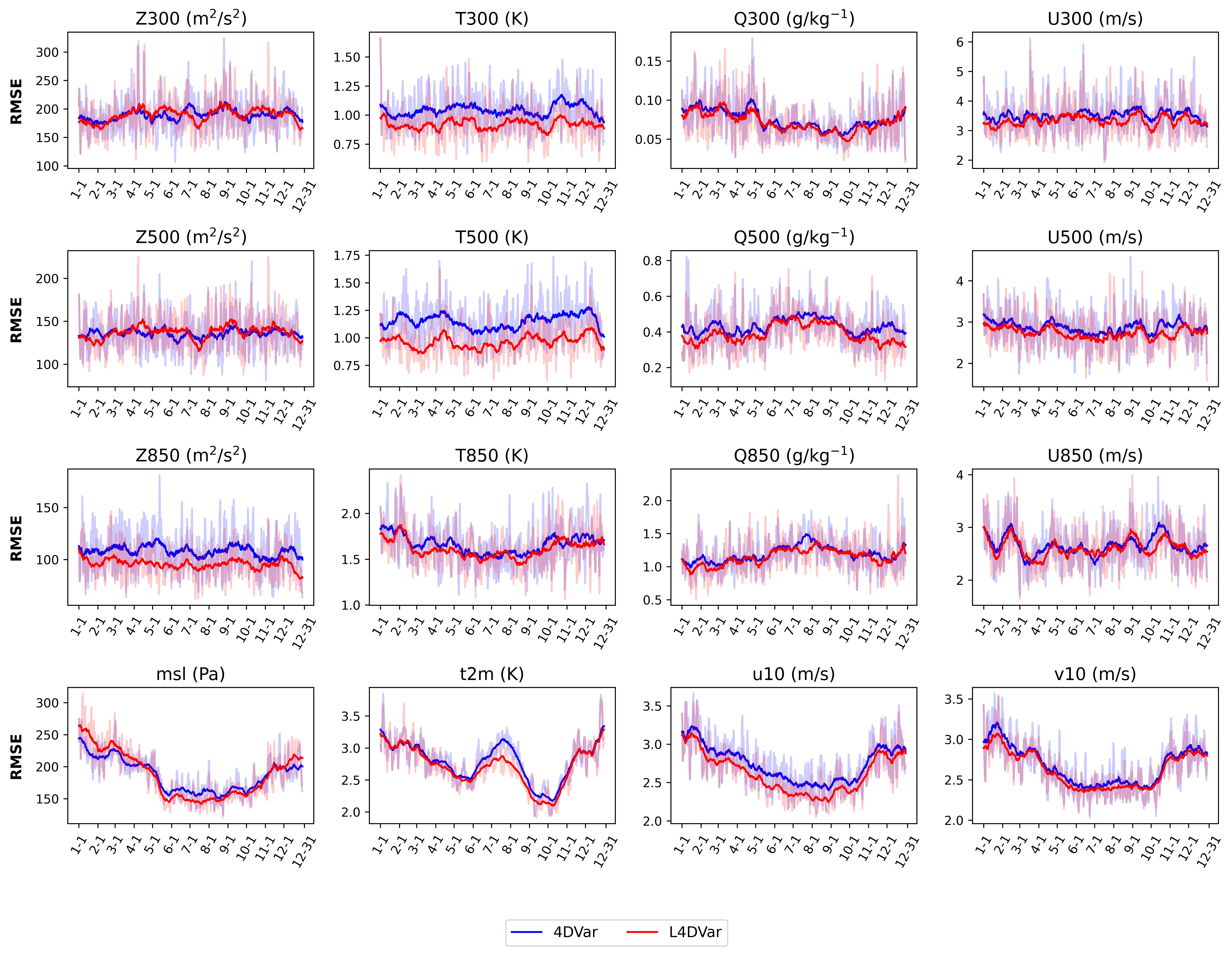}
    \caption{\textbf{Extended analysis performance comparison of 4DVar and L4DVar with real observations.}
    This figure expands on Fig. 2C by presenting RMSE for variables at 300 hPa, 500 hPa, 850 hPa, and the surface, evaluated against observations withheld from assimilation. }
    \label{fig:analysis_extend}
\end{figure}

\begin{figure}[p]
    \centering
    \includegraphics[width=\linewidth]{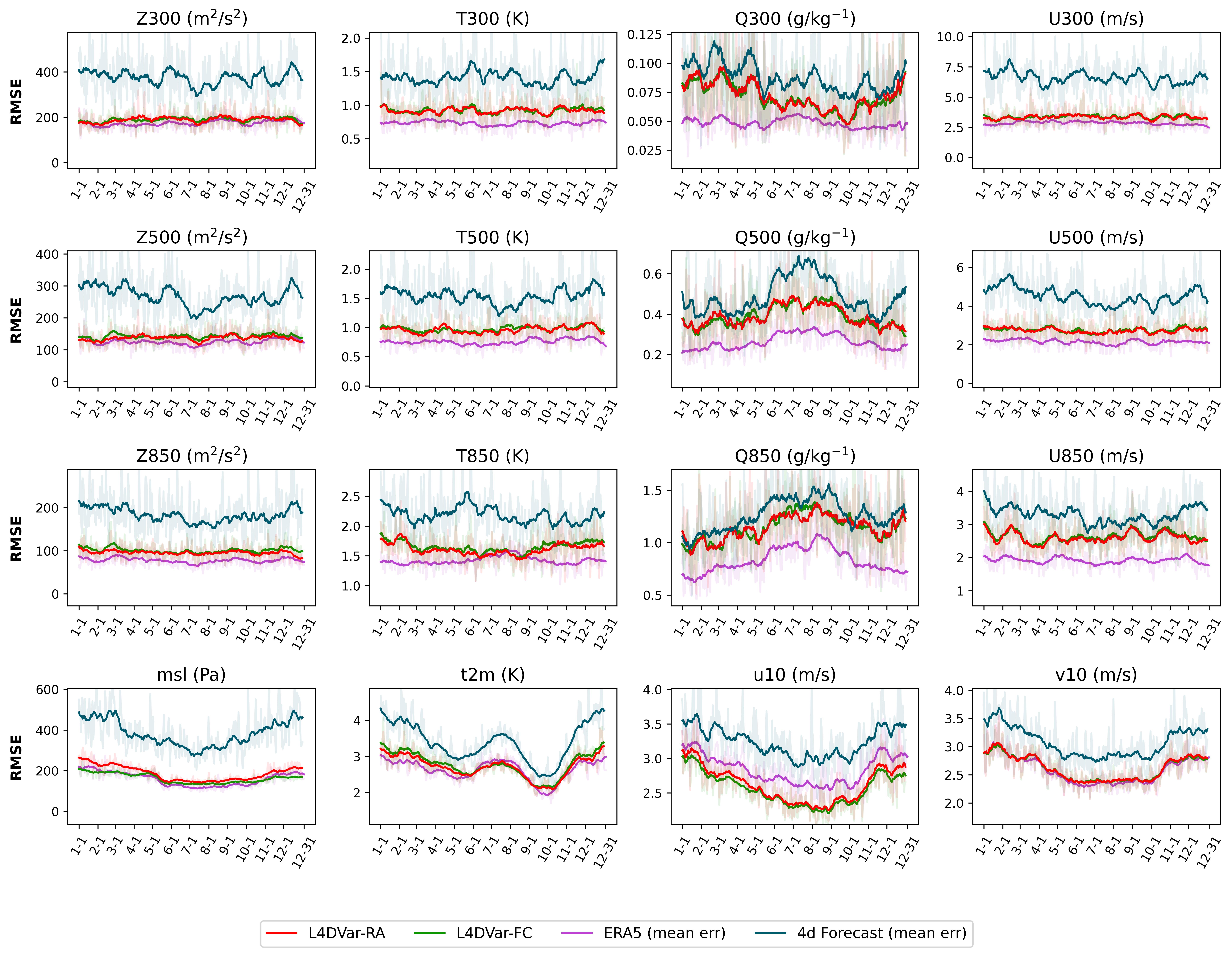}
    \caption{\textbf{Extended analysis performance comparison of L4DVar using AEs trained on forecasts and reanalysis.} This figure expands on Fig. 2D by presenting RMSE for variables at 300 hPa, 500 hPa, 850 hPa, and the surface, evaluated against observations withheld from assimilation. Results are shown for L4DVar using AE trained on either ERA5 reanalysis (L4DVar-RA) or 5-day forecasts (L4DVar-FC). For reference, the RMSE of ERA5 and forecast data at corresponding analysis times is also included to reflect the quality of their training datasets.}
    \label{fig:forecast_analysis_extend}
\end{figure}

\begin{figure}[p]
    \centering
    \includegraphics[width=\linewidth]{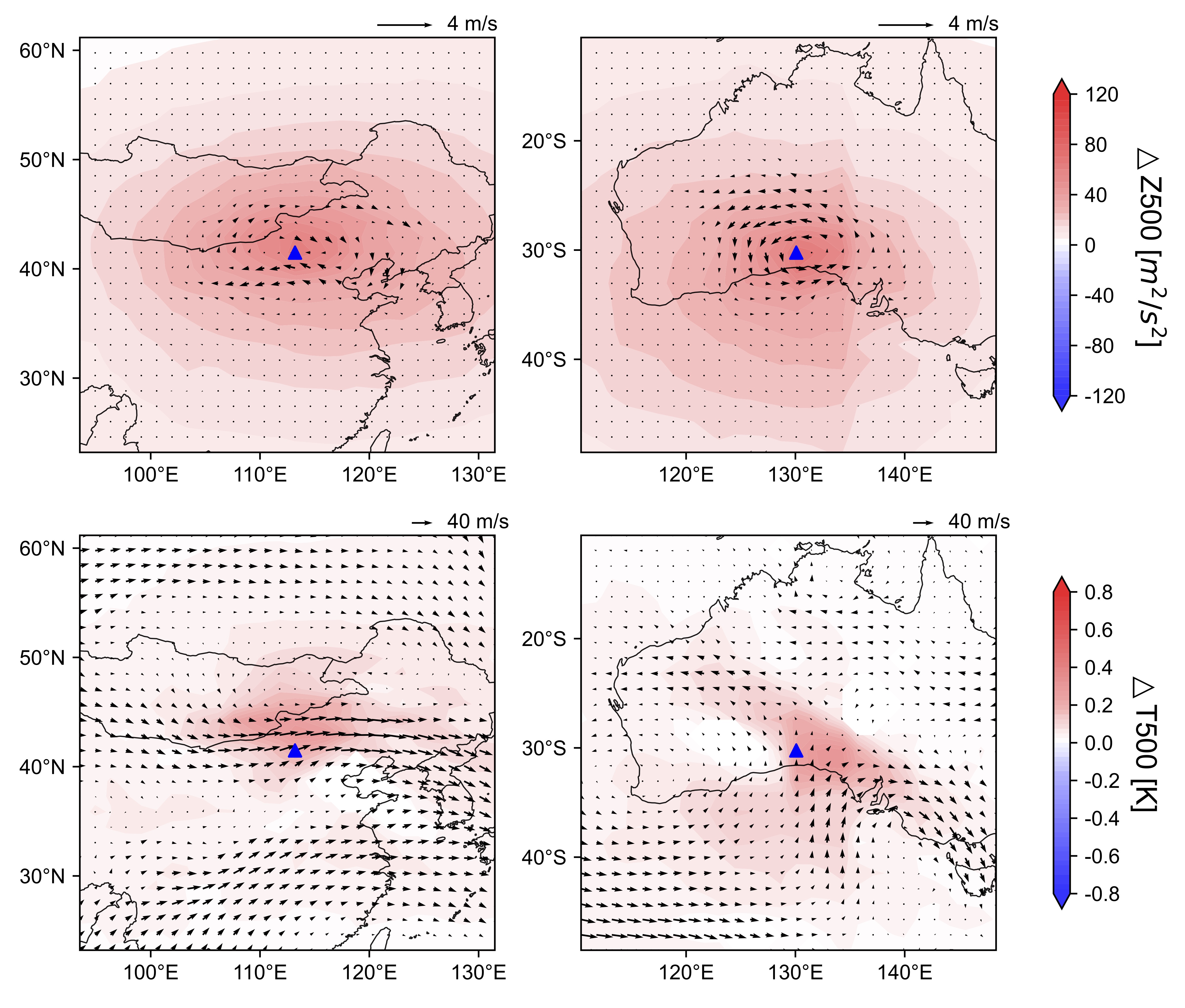}
    \caption{\textbf{Additional single-observation experiment demonstrating physical consistency in LDA.}
    This plot replicates the setup of Fig. 3A, but uses ERA5 reanalysis at 0000 UTC on 1 January 2017 as the background field.}
    \label{fig:single_obs_sample2}
\end{figure}

\begin{figure}[p]
    \centering
    \includegraphics[width=\linewidth]{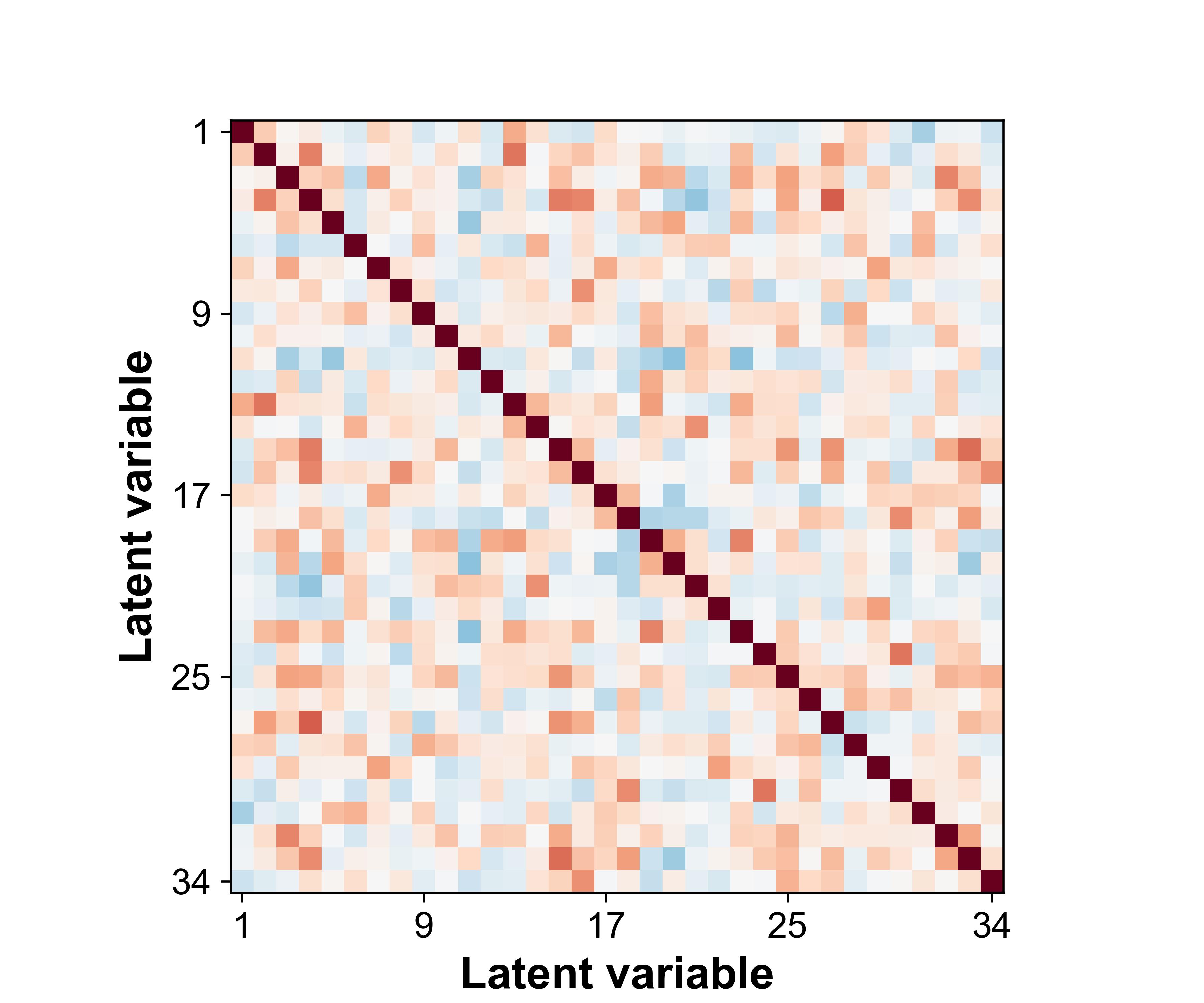}
    \caption{\textbf{Inner product matrix of the row vectors from Fig. 3B in the main text, representing variable-wise response patterns induced by individual latent variable increments.} Low off-diagonal values indicate that different latent variables represent distinct and approximately orthogonal relationships among atmospheric variables.}
    \label{fig:inner_product}
\end{figure}

\begin{figure}[p]
    \centering
    \includegraphics[width=\linewidth]{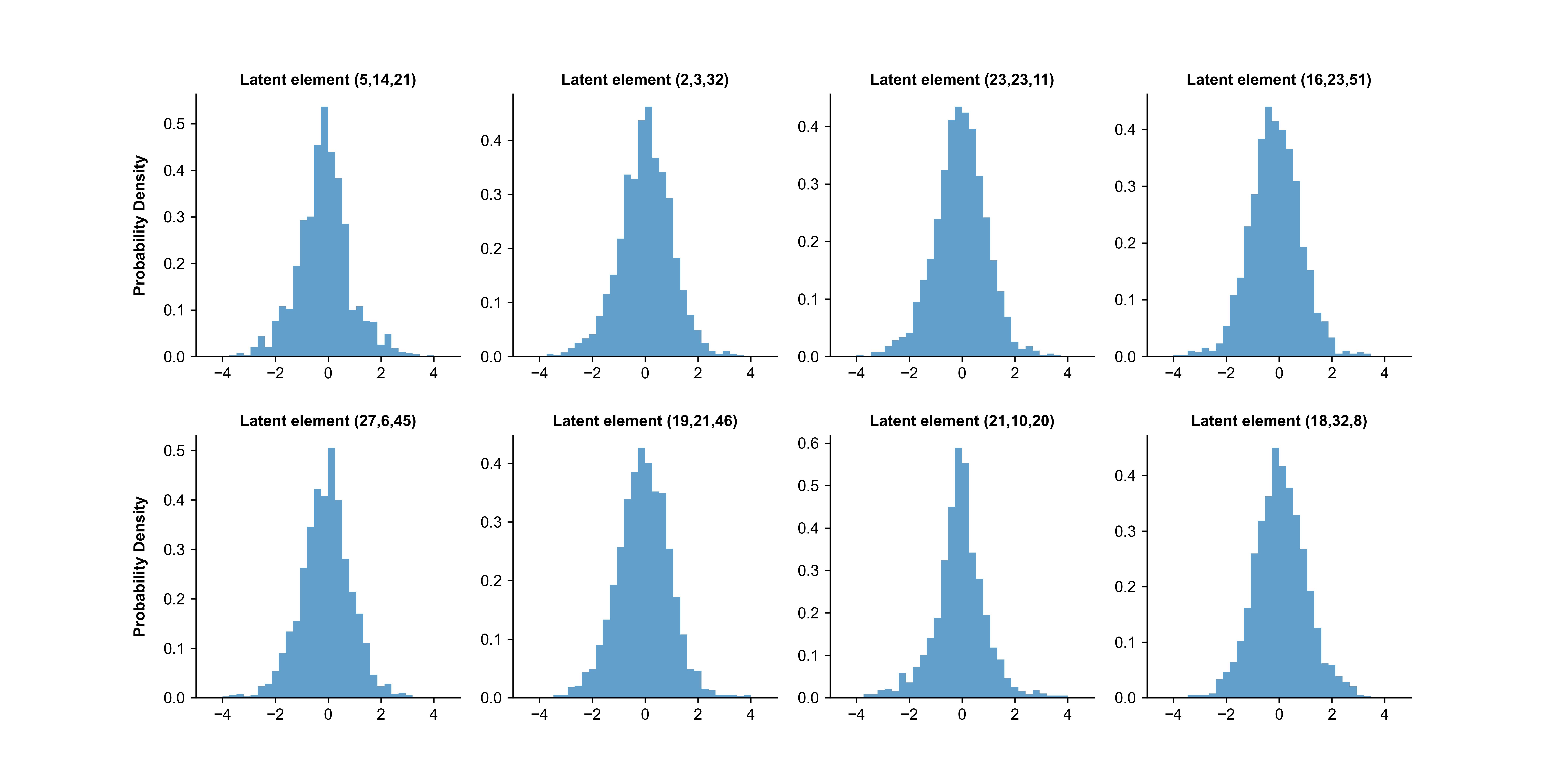}
    \caption{\textbf{Probability distribution of background errors in latent space.} Background errors were estimated using the NMC method and projected into the latent space. Several latent dimensions were randomly selected and their error distributions are shown.}
    \label{fig:Gaussian_distribution}
\end{figure}

\begin{figure}[p]
    \centering
    \includegraphics[width=\linewidth]{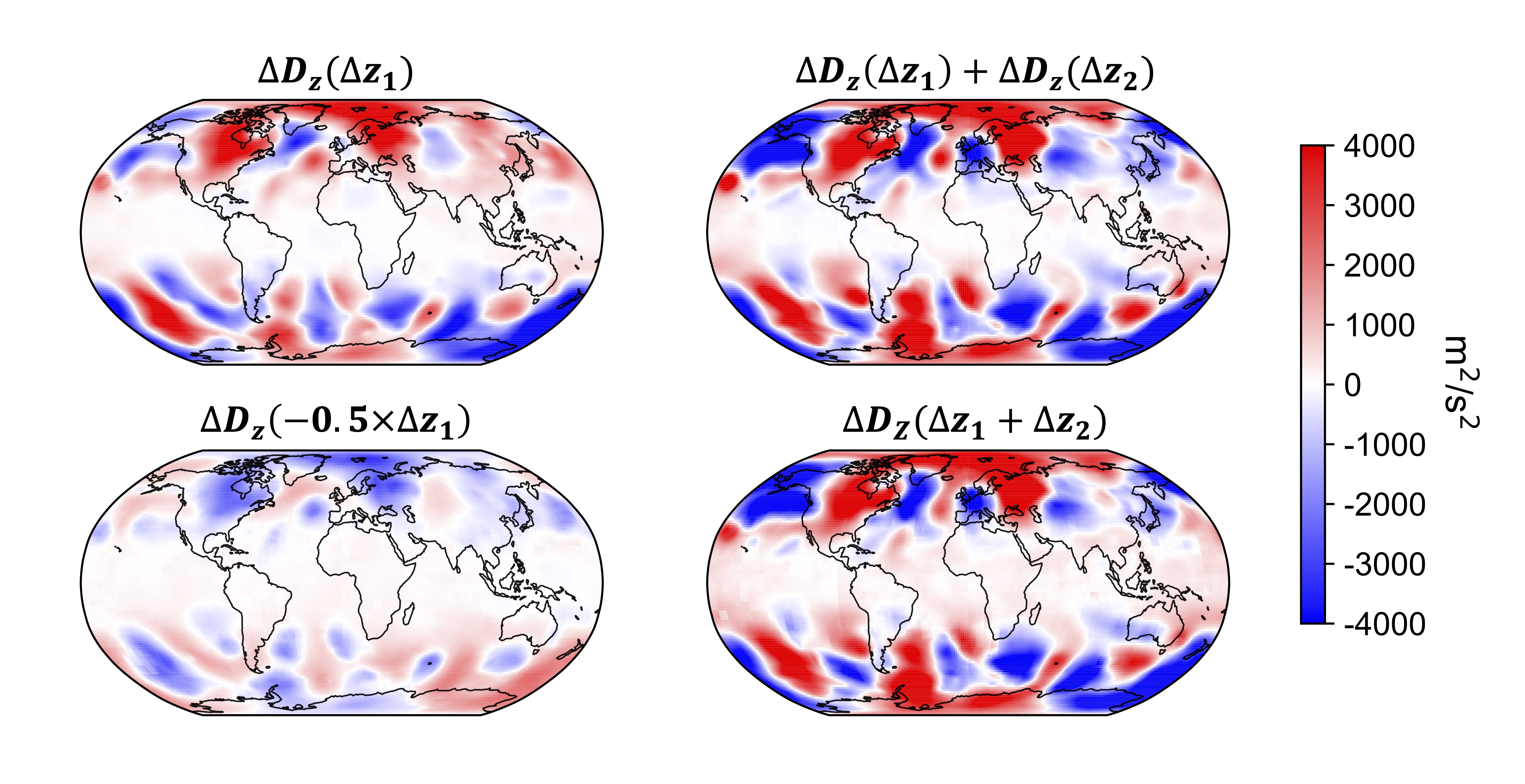}
    \caption{\textbf{Extended example demonstrating approximate affinity of the AE decoder along latent directions representative of
    atmospheric variability.} Here, $\boldsymbol{z}$ denotes the latent state corresponding to the ERA5 reanalysis at 0000 UTC on November 1, 2017. The perturbations $\Delta \boldsymbol{z_1}$ and $\Delta \boldsymbol{z_2}$ represent the latent differences between $\boldsymbol{z}$ and the reanalysis at 0000 UTC on October 1 and December 1, 2017, respectively.}
    \label{fig:affine_sample2}
\end{figure}

\begin{figure}[p]
    \centering
    \includegraphics[width=\linewidth]{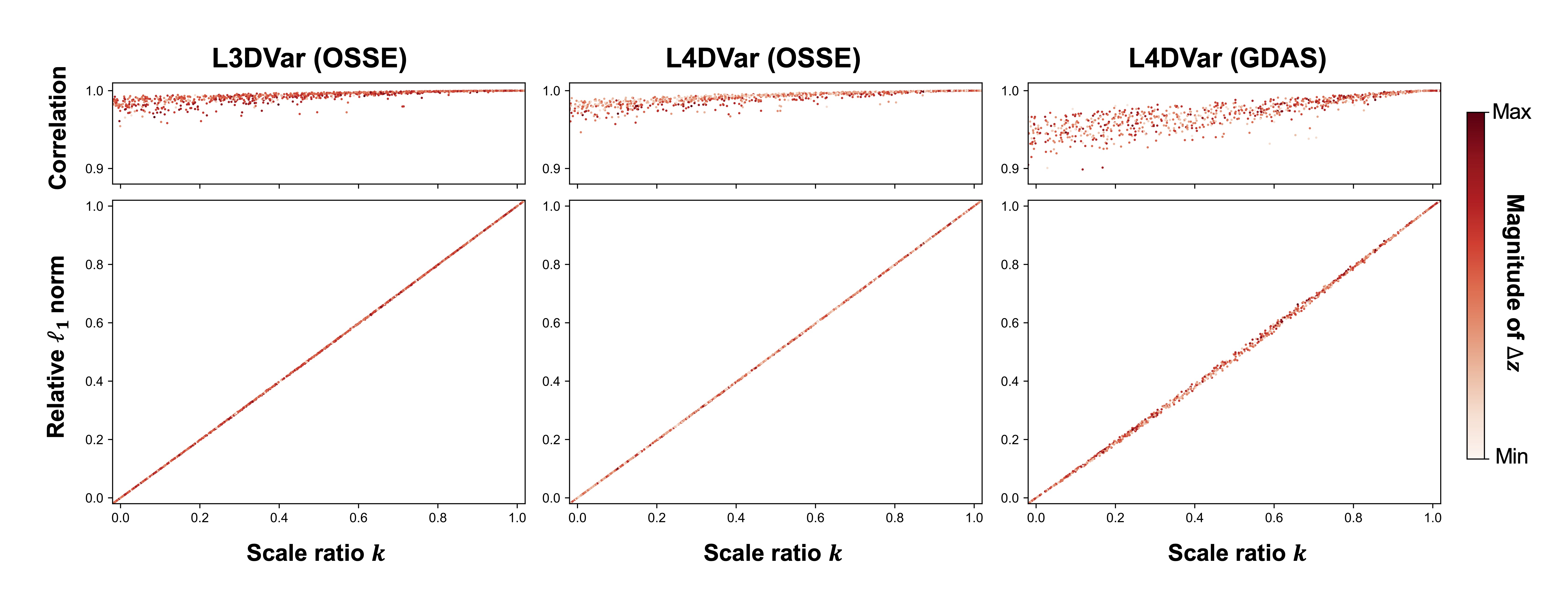}
    \caption{\textbf{Approximate affinity of the AE decoder across LDA experiments.} 
    The affinity of the AE decoder is assessed by comparing the decoded increments $\Delta D_{\boldsymbol{z}}(k \cdot \Delta \boldsymbol{z})$ and $k \cdot \Delta D_{\boldsymbol{z}}(\Delta \boldsymbol{z})$ across varying scale ratios $k$. For each $\Delta \boldsymbol{z}$, 10 different values of $k$ uniformly sampled from the range [0, 1] are tested. The top panels show their correlation coefficients, while the bottom panels depict the relative $\ell_1$ norms. This demonstrates that the decoder exhibits approximate local affinity throughout the LDA process, across both OSSEs and real-observation experiments (Fig. 1A, B in the main text).}
    \label{fig:LDA_decoder_affine}
\end{figure}

\clearpage
\begin{figure}[p]
    \centering
    \includegraphics[width=\linewidth]{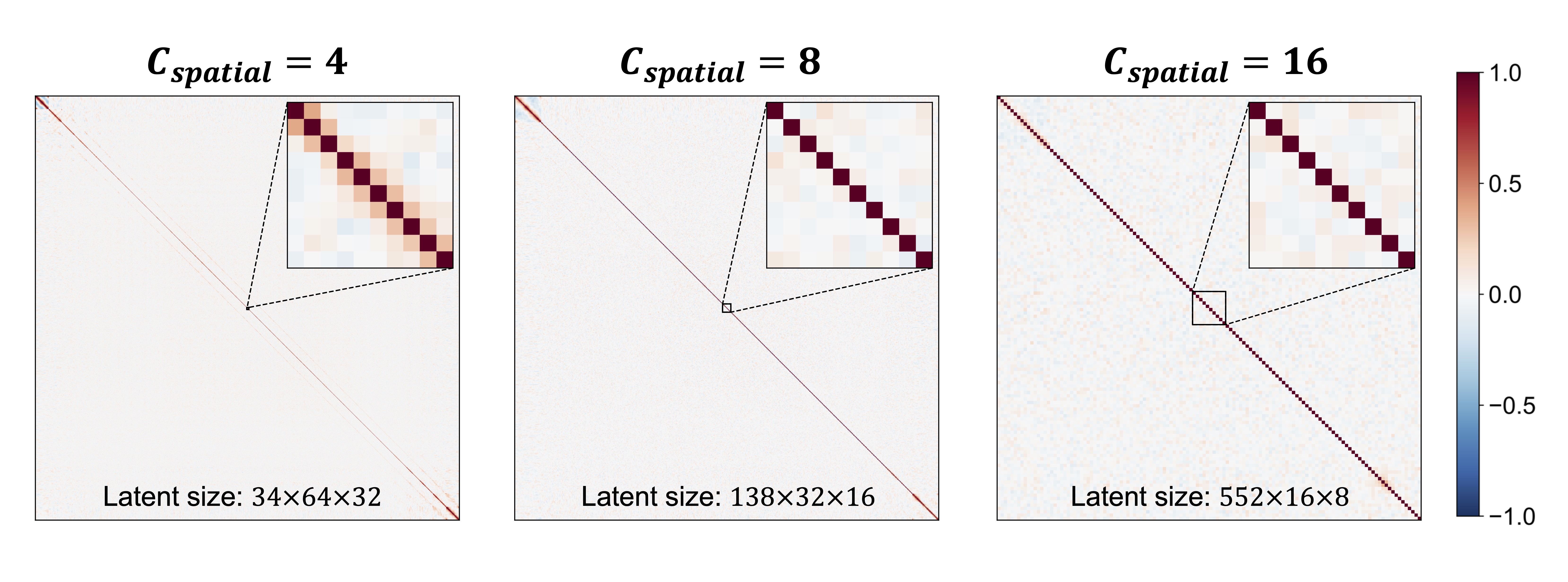}
    \caption{\textbf{Spatial correlation of background errors in latent space under varying spatial compression ratios $C_{\text{spatial}}$.}
    Correlations are computed using the same NMC-generated samples across all settings, with a fixed total latent size but varying $C_{\text{spatial}}$. Results show that residual spatial correlations persist when the $C_{\text{spatial}}$ is 4, but these correlations largely vanish and saturate as the $C_{\text{spatial}}$ increases to 8.}
    \label{fig:decouple_effect}
\end{figure}

\begin{figure}[p]
    \centering
    \includegraphics[width=\linewidth]{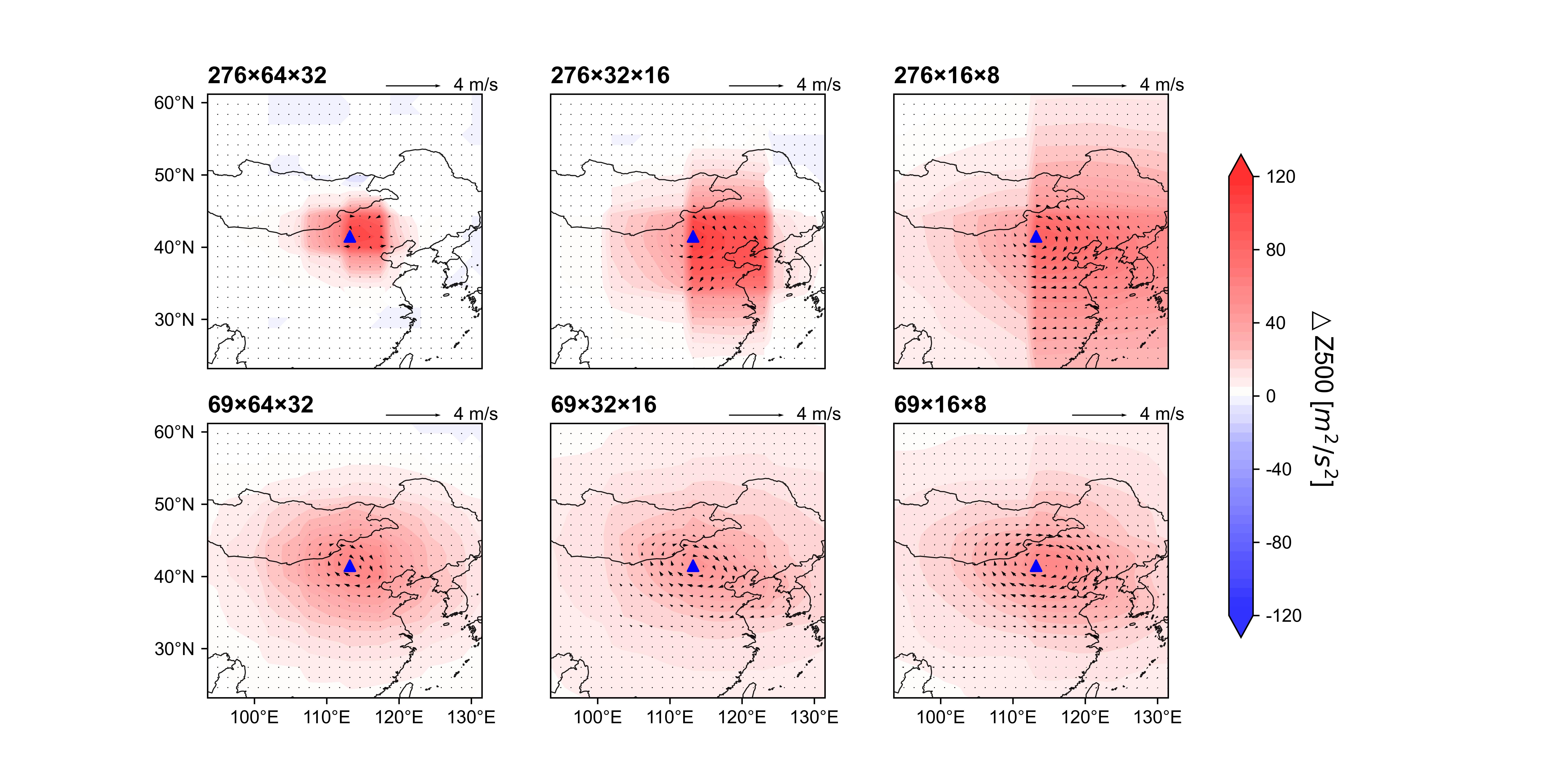}
    \caption{\textbf{Illustration of patch-like increments in single-observation experiments under varying latent dimensionalities.}
    Analysis increments are computed using LDA with AEs of different latent sizes, in response to a 200 m$^2$/s$^2$ perturbation applied to geopotential height at 500 hPa (marked by the blue triangle). When the variable dimension (i.e., the number of channels) of the latent space exceeds that of the model space (as in the top panel), the resulting increments exhibit patch-like structures. These patterns remain physically consistent, with stronger geopotential height anomalies accompanied by stronger geostrophic wind responses. An exception occurs for the AE with a latent size of 276×64×32, where insufficient compression prevents the latent space from preserving balance properties.}
    \label{fig:patch_effect}
\end{figure}






\clearpage
\begin{table}
    \centering
    \caption{\textbf{Statistical assessment of the approximate affinity of the AE decoder along latent directions representative of atmospheric variability.}
    We assess whether the AE decoder exhibits affine behavior in response to latent perturbations around atmospheric latent states $\boldsymbol{z}$, by evaluating whether the decoded increment $\Delta D_{\boldsymbol{z}}(\Delta \boldsymbol{z})$ (defined as $D(\boldsymbol{z}+\Delta \boldsymbol{z})-D(\boldsymbol{z})$) satisfies two key properties of affine transformations: homogeneity and additivity. 
    To this end, 10{,}000 ERA5 reanalysis states are randomly sampled as $\boldsymbol{z}$, and latent increments $\Delta \boldsymbol{z}$ are constructed by differencing with other randomly selected samples. 
    For the homogeneity test, $k$ is uniformly drawn from $[-2, 2]$. 
    The subscript $i$ indicates a perturbation applied only to the $i$th latent variable.
    All reported metrics are averaged over the 10{,}000 test cases. 
    Results show that the AE decoder behaves approximately affinely along these representative directions, which dominate the variability of atmospheric states.}
    \label{tab:decoder_affinity}

    \begin{tabular}{lcc}
        \hline
        & \textbf{Homogeneity} & \textbf{Additivity} \\
        \hline
        Compared fields 
        & $\Delta D_{\boldsymbol{z}}(k \cdot \Delta \boldsymbol{z})$ \textbf{vs.} $k \cdot \Delta D_{\boldsymbol{z}}(\Delta \boldsymbol{z})$ 
        & $\Delta D_{\boldsymbol{z}}(\boldsymbol{z}_1 + \boldsymbol{z}_2)$ \textbf{vs.} $\Delta D_{\boldsymbol{z}}(\boldsymbol{z}_1)+\Delta D_{\boldsymbol{z}}(\boldsymbol{z}_2)$ \\
        Absolute Correlation & $0.91$ & $0.98$ \\
        Relative $\ell_1$ norm & $1.04$ & $0.99$ \\
        \hline
    \end{tabular}
\end{table}

\end{document}